\documentclass[a4paper,fleqn,usenatbib]{mnras}

\usepackage{mathptmx}

\usepackage{ae,aecompl}

\usepackage{graphicx}	
\usepackage{amsmath}	
\usepackage{amssymb}	

\usepackage{float}

\usepackage{color,epsfig} 
\usepackage[utf8]{inputenc} 

\usepackage{multirow}

\usepackage[flushleft]{threeparttable}
\makeatletter
\newlength{\abovecaptionskip}%
\makeatother

\newcommand{\be}{\begin{equation}}
\newcommand{\ee}{\end{equation}}

\providecommand{\comment}[1]{} 

\newcommand*\diff{\mathop{}\!\mathrm{d}}
\newcommand{\vect}[1]{\mathbfit{#1}}
\newcommand{\tens}[1]{\mathbb{#1}}

\newcommand{\mstar}{M_{\star}}
\newcommand{\rstar}{R_{\star}}

\newcommand{\rp}{R_{\rm p}}
\newcommand{\rt}{R_{\rm t}}

\newcommand{\rint}{R_{\rm int}}

\newcommand{\vn}{v_{\rm n}}
\newcommand{\ve}{v_{\rm e}}

\newcommand{\mh}{M_{\rm h}}

\newcommand{\tmin}{t_{\rm min}}
\newcommand{\amin}{a_{\rm min}}

\newcommand{\ex}{\vect{e}_{\rm x}}
\newcommand{\ey}{\vect{e}_{\rm y}}
\newcommand{\ez}{\vect{e}_{\rm z}}

\newcommand{\rg}{R_{\rm g}}

\newcommand{\mdotp}{\dot{M}_{\rm p}}
\newcommand{\mdotedd}{\dot{M}_{\rm Edd}}

\def\m6{\, M_6}

\def\a3{\, a_3}

\def\msun{\, \mathrm{M}_{\hbox{$\odot$}}}
\def\rsun{\, \mathrm{R}_{\hbox{$\odot$}}}
\def\gcm3{\, \rm g \, cm^{-3}}
\def\gpercm2{\, \rm g \, cm^{-2}}
\def\ergcm3{\, \rm erg \, cm^{-3}}
\def\ergcm3s{\, \rm erg \, cm^{-3} \, s^{-1}}

\def\days{\, \rm d}

\def\kelvin{\, \rm K}

\def\ergpers{\, \rm erg\, s^{-1}}
\def\msunperyr{\, \rm \mathrm{M}_{\hbox{$\odot$}} \, {yr}^{-1}}
\def\cm2g{\, \rm cm^{2} \, g^{-1}}

\definecolor{orange}{RGB}{236,133,40}

\title[First light from TDEs]{First light from tidal disruption events}

\author[C. Bonnerot, W. Lu and P. F. Hopkins]{Clément Bonnerot$^{1}$\thanks{E-mail: bonnerot@tapir.caltech.edu}, Wenbin Lu$^{1}$\thanks{E-mail: wenbinlu@caltech.edu} and Philip F. Hopkins$^{1}$
\\
$^{1}$TAPIR, Mailcode 350-17, California Institute of Technology, Pasadena, CA 91125, USA\\
}
\date{Accepted XXX. Received YYY; in original form ZZZ}

\pubyear{2020}

\begin{document}
\label{firstpage}
\pagerange{\pageref{firstpage}--\pageref{lastpage}}
\maketitle

\begin{abstract}

When a star comes too close to a supermassive black hole, it gets torn apart by strong tidal forces in a tidal disruption event, or TDE. Half of the elongated stream of debris comes back to the stellar pericenter where relativistic apsidal precession induces a self-crossing shock. As a result, the gas gets launched into an outflow that can experience additional interactions, leading to the formation of an accretion disc. We carry out the first radiation-hydrodynamics simulations of this process, making use of the same injection procedure to treat the self-crossing shock as in our previous adiabatic study \citep{bonnerot2020-realistic}. Two sets of realistic parameters of the problem are considered that correspond to different strengths of this initial interaction. In both cases, we find that the injected matter has its trajectories promptly circularized by secondary shocks taking place near the black hole. However, the generated internal energy efficiently diffuses away in the form of radiation, which results in a thin vertical profile of the formed disc. The diffusing photons promptly irradiate the surrounding debris until they emerge with a bolometric luminosity of $L\approx 10^{44} \ergpers$. Towards the self-crossing shock, diffusion is however slowed that results in a shallower luminosity increase, with a  potentially significant component in the optical band. Matter launched to large distances continuously gains energy through radiation pressure, which can cause a significant fraction to become unbound. This work provides direct insight into the origin of the early emission from TDEs, which is accessed by a rapidly increasing number of observations.

\end{abstract}




\section{Introduction}
\label{introduction}

Stars occasionally get close enough to the supermassive black hole at the center of massive galaxies to be torn apart by strong tidal forces \citep{rees1988}, a phenomenon referred to as tidal disruption event (TDE). Following this encounter, the stellar debris evolves into an elongated stream, half of which remains bound and falls back towards the black hole after an entire revolution around it. This return of the debris to the disruption site results in a luminous multi-wavelength emission likely powered by a combination of shocks and accretion from a nascent disc. This signal represents a unique probe of the black hole and the various physical processes occurring in its vicinity, which include the stellar interactions leading to the plunging orbit of the victim star and the physics of super-Eddington accretion.

Thanks to the advent of several high-cadence surveys, the sample of detected TDEs is rapidly increasing both in size and quality, a trend that should continue in the future at an even faster pace with new facilities such as eROSITA \citep{khabibullin2014} and LSST \citep{Bricman2020}. In particular, multiple well-sampled observations in the optical and UV bands have now managed to catch the early rise to the peak of the lightcurve, which probes the first interactions experienced by the debris as they fall back near the black hole \citep{gezari2012,chornock2014,arcavi2014,blagorodnova2017,leloudas2019,holoien2019,holoien2019-18kh,nicholl2019,van_velzen2019,holoien2020,nicholl2020}. This phase is often associated with signs of outflow evidenced by the detection of radio emission and blue-shifted atomic lines \citep[e.g.][]{nicholl2020}. Strong Bowen fluorescence lines have also been detected \citep[e.g.][]{blagorodnova2019,leloudas2019} whose formation necessitates the presence of extreme-ultraviolet radiation likely produced by prompt gas accretion. This scenario is additionally supported in some events by the early detection of a double-peaked emission line \citep{hung2020,short2020}, potentially originating from a nearly circular accretion disc.

Optimally exploiting this wealth of observational data requires to improve our theoretical understanding of these events. Many analytical and numerical investigations of the stellar disruption \citep{carter1982,evans1989,lodato2009,brassart2010,guillochon2013,stone2013} and stream evolution \citep{kochanek1994,coughlin2015-variability,coughlin2016-structure,steinberg2019} have been carried out that evaluated the gas evolution during these early phases to a now satisfactory level of precision. However, the fate of this gas after it comes back near the black hole to form an accretion disc remains much more uncertain. This is both due to a higher complexity and the very high computational cost required to study it from realistic initial conditions \citetext{see \citealt{bonnerot2020-review} for a recent review}. Current works nevertheless suggest that this process is initiated by a self-crossing shock caused by an intersection between the part of the stream moving away from the black hole after passing pericenter and that still infalling. In order to reduce numerical demands, most global simulations have been carried out using artificially bound stellar orbits \citep{hayasaki2013,bonnerot2016-circ,hayasaki2016-spin,sadowski2016,liptai2019-spin} or intermediate-mass black holes \citep{ramirez-ruiz2009,guillochon2014-10jh,shiokawa2015}. By modelling the self-crossing shock through an injection of outflowing gas inside the computational domain, \cite{bonnerot2020-realistic} were able to follow this process for realistic parameters of the problem, that is a supermassive black hole and parabolic stellar trajectory. More recently, this has also been achieved globally \citep{andalman2020} using a GPU-accelerated code, although the authors had to consider a less likely deep encounter and evolve the gas for a shorter time past pericenter to limit computational overhead.

The above numerical investigations find that kinetic energy dissipation through the self-crossing shock and subsequent interactions cause the gas trajectories to become more circular until it settles into an accretion disc. Most of them make the assumption of gas adiabaticity to treat its thermodynamical evolution, which results in the formation of a very thick torus that retains all the internal energy injected by the circularizing shocks. This is legitimate when evolving the gas through the initial self-crossing shock where the large optical depths of the original stream prevents efficient cooling \citep{kim1999,jiang2016}. However, gas expansion at later times decreases its optical thickness such that radiative diffusion may play a significant role \citep{piran2015-circ,hayasaki2016-spin}. The extreme situation where all the internal energy is instantaneously lost from the system by radiation has been studied through isothermal simulations for the simplified case of bound stars \citep{bonnerot2016-circ,hayasaki2016-spin,liptai2019-spin}. Due to reduced hydrostatic support against gravity, they find that the formed disc becomes much thinner and less extended than in the adiabatic cases. The radiation produced has to diffuse through the surrounding stellar debris until it reaches the scattering photosphere, where photons can emerge from the gas. During this outward transport, radiation can experience adiabatic losses as well as a shift of its mean energy to lower values due to absorption and re-emission, potentially accounting for the signal detected from TDEs at optical wavelengths \citep{guillochon2014-10jh,metzger2016,roth2016}. Taking the above effects into account self-consistently requires to study the coupled evolution of gas and radiation as the debris stream returns near the black hole.

In this paper, we carry out the first radiation-hydrodynamics simulations of disc formation using a meshless numerical approach. Realistic parameters of the problem are considered by using the same strategy as \cite{bonnerot2020-realistic} that consists in treating the self-crossing shock through an injection of outflowing gas according to the properties obtained from a local numerical study of this initial interaction \citep{lu2020}. Radiation is produced though free-free emission at shocks and then evolved using the flux-limited diffusion (FLD) technique, which is appropriate given the large optical thickness of the stellar debris. 
Its impact on the gas dynamics through radiation pressure as the disc forms is also accounted for as well as the associated adiabatic losses. Through the amount of radiation leaving the system, we are able to evaluate the emerging luminosity, which represents the first\footnote{Prior to the return of the debris to the black hole, emission may occur due to hydrogen recombination within the stream \citep{kasen2010}. However, the luminosity of this signal is of $L_{\rm rec} \lesssim 10^{41} \ergpers$, which has so far not been detected and is much fainter than that following the formation of an accretion disc. As it passes at pericenter, the stream also experiences dissipation due to strong vertical compression. However, the resulting internal energy injection is much lower than that associated with disc formation and the photons produced are more susceptible to adiabatic losses due to the stream compactness.} detectable signal from a TDE.

The paper is presented as follows. We describe in Section \ref{sec:meshless} the meshless simulations used by explaining our initial conditions as well as the numerical treatment for the coupled evolution of gas and radiation. The results are presented in Section \ref{sec:results} starting from radiative diffusion from the secondary shocks experienced by the debris and the resulting formation of an accretion disc. We then study the transport of radiation through the gas until it emerges from the system as well as the properties of the outflowing matter. A discussion is included in Section \ref{sec:discussion} and we summarize our findings in Section \ref{sec:summary}.

\section{Meshless simulations}
\label{sec:meshless}

We study the process of disc formation through radiation-hydrodynamics simulations by following the gas evolution after it has passed through the self-crossing shock. This is achieved by continuously injecting matter into the computational domain, a technique presented in our previous investigation \citep{bonnerot2020-realistic} that assumed an adiabatic evolution for the gas. In the present paper, we build on this earlier study by incorporating an improved physical treatment of radiation as well as expanding the exploration of the parameter space.

\subsection{Initial conditions}

\label{sec:initial}

The star encounters the black hole on a near-parabolic orbit that enters the tidal radius 
\be
\rt = \rstar \left(\frac{\mh}{\mstar}\right)^{1/3},
\label{eq:tidal-radius}
\ee
where $\mh$ is the black hole mass while $\mstar$ and $\rstar$ are the stellar mass and radius, respectively. The pericenter of this trajectory is typically parametrized by $\rp = \rt/\beta$, where $\beta$ is the penetration factor that determines the depth of the disruption. During this process, the stellar elements get imparted a spread in orbital energy $\Delta \epsilon = G \mh \rstar / \rt^2$ \citep{stone2013} that results in their evolution into an elongated stream, roughly half of which gets ejected to infinity while the rest becomes bound and returns to the disruption site after an entire revolution around the black hole. The most bound debris has the smallest semi-major axis given by
\be
\amin = \frac{G \mh}{2 \Delta \epsilon} = \frac{\rstar}{2} \left(\frac{\mh}{\mstar} \right)^{1/3},
\label{eq:semi-major-axis}
\ee
and is the earliest to fall back to pericenter after an orbital period of 
\be
\tmin = 2^{-1/2} \pi \left(\frac{G \mstar}{\rstar^3}\right)^{-1/2} \left(\frac{\mh}{\mstar}\right)^{1/2},
\ee
according to Kepler's third law. Assuming a flat orbital energy distribution through the debris, the rate at which mass falls back is given by \citep{rees1988,phinney1989}
\be
\dot{M}_{\rm fb} = \dot{M}_{\rm p} \left(1+\frac{t}{\tmin}\right)^{-5/3},
\label{eq:fallback_rate}
\ee
where the time is set to $t=0$ at the peak value given by 
\be
\dot{M}_{\rm p} = \frac{\mstar}{3 \tmin}.
\ee
For simplicity, we adopt here this analytical form of the fallback rate that always decreases with time. However, hydrodynamical simulations of the disruption \citep{lodato2009,guillochon2013} find that it additionally features an initial rise due to the lower density of the most bound material originating from the stellar outer layers.

During its pericenter passage, the stream experiences relativistic apsidal precession that causes a self-crossing shock between the part moving away from the black hole and that still approaching it. This collision takes place at an intersection radius contained between the pericenter and the apocenter of the most bound debris, that is $\rp \lesssim \rint \lesssim 2 \amin$. It decreases for more relativistic encounters whose pericenter is close to the gravitational radius of the black hole given by $\rg = G \mh /c^2$ due to stronger apsidal precession \citep{dai2015,bonnerot2017-stream}. Local simulations of the self-crossing shock assuming identical and thin stream components\footnote{Although the initial conditions adopted in these local studies are consistent with our current understanding of the previous stream evolution, the exact properties of the colliding streams are still uncertain that may affect the outcome of the self-crossing shock. The infalling component likely retains a thin width due to its previous confinement by self-gravity, although it may additionally expand due to hydrogen recombination \citep{kasen2010,coughlin2016-structure} or if the stream is highly magnetized \citep{guillochon2017-magnetic,bonnerot2017-magnetic}. However, the outgoing part of the stream may be significantly thicker due to heating through shocks taking place during its pericenter passage, as we further discuss in Section \ref{sec:self}.} find that the collision is close to adiabatic that results in the formation of a quasi-spherical outflow \citep{jiang2016}. As described by \citet{lu2020}, the dynamics of the outflowing gas can then be accurately described by an expansion velocity $\vect{v}_{\rm e}$ directed radially away from the intersection point. As expected from energy conservation, this speed is equal in magnitude to the radial component of the colliding streams velocity, whose associated kinetic energy gets dissipated during the shock. In addition, this expanding gas possesses a net speed $\vect{v}_{\rm n}$ tangential to the black hole due to the common motion of the colliding stream components along this direction. In the frame co-moving at this tangential speed, this work also finds that the outflow density is azimuthally symmetric around the common axis along which the streams move before the interaction. This density profile is however enhanced by a factor of a few along the equatorial plane with a polar dependence of the density given by
\be
\bar{\rho}_{\rm sc}(\vartheta) = -0.08436 x^4 + 0.2434 x^3 - 0.2017 x^2 + 0.1103 x + 0.01051,
\label{eq:density-profile}
\ee
where $x \equiv \min(\vartheta,\pi-\vartheta)$ and $\vartheta$ denotes the polar angle, with the normalization being set by $2\pi\int_0^{\pi} \bar{\rho}_{\rm sc}(\vartheta) \sin \vartheta \diff \vartheta = 1$. We refer to figure 1 of our previous study \citep{bonnerot2020-realistic} for a sketch illustrating the configuration described above. The rate at which matter passes through the self-crossing shock is set to
\be
\dot{M}_{\rm sc} = \dot{M}_{\rm fb},
\label{eq:outflow_rate}
\ee
as given by the fallback rate of equation \eqref{eq:fallback_rate}. While we assume that this interaction is entirely continuous, it is in fact intermittent since the outgoing stream component needs to replenish itself after all of this gas has passed through the intersection point. This caveat is discussed in Section \ref{sec:self}, where we identify possible consequences of this intermittency on the dynamics and associated radiation.

We carry out two simulations for different sets of parameters of the problem that correspond to different regimes for the self-crossing shock, as listed in Table \ref{tab:parameters}. In simulation A, we adopt the same as in our previous study \citep{bonnerot2020-realistic}, that is a star with mass $\mstar = 0.5\msun$ and radius $\rstar = 0.46 \rsun$ disrupted by a black hole of mass $\mh = 2.5 \times 10^6 \msun$ with a penetration factor $\beta = 1$. The corresponding period of the most bound debris is $\tmin = 40 \days$, which leads to a fallback rate of  $\mdotp = 1.5 \msunperyr \approx 23 \,\mdotedd$. Here, the Eddington accretion rate is calculated as $\dot{M}_{\rm Edd} = L_{\rm Edd}/(\eta c^2)$, where 
$L_{\rm Edd} = 4 \pi G \mh c / \kappa_{\rm s} \approx 4 \times 10^{44}\ergpers$ denotes the Eddington luminosity for the electron-scattering opacity $\kappa_{\rm s} = 0.34 \cm2g$ and we adopt an accretion efficiency of $\eta = 0.1$. The pericenter distance is significantly relativistic with $\rp \approx 15 \rg$ and the self-crossing shock takes place at $\rint \approx 25 \rp$ where the gas is launched with expansion and net speeds $\ve = 0.065 c$ and $\vn = 0.0152 c$, according to the geodesics calculation carried out by \cite{lu2020}. About $33 \%$ of this outflowing gas is initially unbound from the black hole, moving in the direction of its net speed. Most of the remaining bound matter therefore rotates on average in the direction opposite to that of the original star. In simulation B, the star is slightly more massive with $\mstar = \msun$ and $\rstar = 0.89 \rsun$, while the black hole mass is reduced to $\mh = 10^6 \msun$ with the same penetration factor $\beta = 1$ as before. For these parameters, the most bound debris has a shorter period than for simulation A of $\tmin = 34 \days$. Along with the increased stellar mass, this leads to a larger fallback rate of $\mdotp = 3.5 \msunperyr \approx 136 \,\mdotedd$ with an associated Eddington luminosity of $L_{\rm Edd} \approx 10^{44}\ergpers$. Importantly, the encounter is also less relativistic than in the other simulation with a pericenter $\rp = 42 \rg$, which leads to a larger intersection radius of $\rint = 72 \rp \approx 1.4 \amin$. Accordingly, the velocity components of the gas passing through the self-crossing shock are reduced to $\ve = 0.013 c$ and $\vn = 0.00306 c$. This implies that all this outflowing gas remains bound to the black hole, which differs from simulation A where a fraction of this gas gets unbound.

\begin{table*}
\begin{threeparttable}
\centering
\caption{Parameters considered in simulations A and B}
\begin{tabular}{|c|c|c|c|c|c|c|c|c|c|}
\hline
 & $\mh$ & $\mstar$ & $\rstar$ & $\beta$ & $\rp/\rg$ & $\rint$\tnote{a} & $\ve$\tnote{a} & $\vn$\tnote{a} & $L_{\rm Edd}$ \\
\hline
Simulation A & $2.5 \times 10^6 \msun$ & $0.5 \msun$ & $0.46 \rsun$ & 1 & 15 & $25 \rp$ & $0.065 c$ & $0.0152 c$ & $4 \times 10^{44} \ergpers$  \\
Simulation B & $10^6 \msun$ & $1 \msun$  & $0.89 \rsun$ & 1 & 42 & $72 \rp \approx 1.4 \amin$  & $0.013 c$ & $0.00306 c$ & $10^{44} \ergpers$ \\
\hline
\end{tabular}
\begin{tablenotes}
\item[a] The values of the intersection radius, expansion and net speeds of the outflow launched from the self-crossing shock are obtained from the other parameters by integrating the general-relativistic trajectory of the returning stream as it precesses around the black hole \citep{lu2020}. 
\end{tablenotes}
\label{tab:parameters}
\end{threeparttable}
\end{table*}

\subsection{Numerical method}

\label{sec:method}

The simulations are carried out with the code
\textsc{gizmo} \citep{hopkins2015} making use of the meshless-finite-mass method that consists in dividing the gas distribution into particles of fixed mass. Like in our previous study \citep{bonnerot2020-realistic}, the self-crossing shock is modelled by continuously injecting gas inside the computational domain. As illustrated in figure 1 of this earlier paper, this injection is made from a small sphere centered on the intersection point. At each timestep $\Delta t$, a number of particles $\Delta N = \left[ \dot{M}_{\rm sc} \Delta t/M_{\rm p} \right]$ is injected where the brackets represent the nearest integer function and the outflow rate is set by equation \eqref{eq:outflow_rate}. The particle mass is set to $M_{\rm p} = 10^{-8} \msun$, for which the resolution study made in \citet{bonnerot2020-realistic} indicates that the simulation has reached an acceptable level of numerical convergence.\footnote{Although this convergence study strictly speaking only applies to simulation A that uses the same parameter as in our earlier work, the similar hydrodynamics seen in simulation B implies that sufficient resolution is achieved by using the same particle mass.} From the start of the self-crossing shock at $t=0$, we evolve the system until $t / \tmin = 0.6$ when the mass of injected debris has reached $27 \%$ of the bound part of the stream, which corresponds to a number of particles of about $N = 6.75 \times 10^6$ and $N = 1.35 \times 10^7$ at the end for simulations A and B, respectively.

\begin{figure*}
\centering
\includegraphics[width=\textwidth]{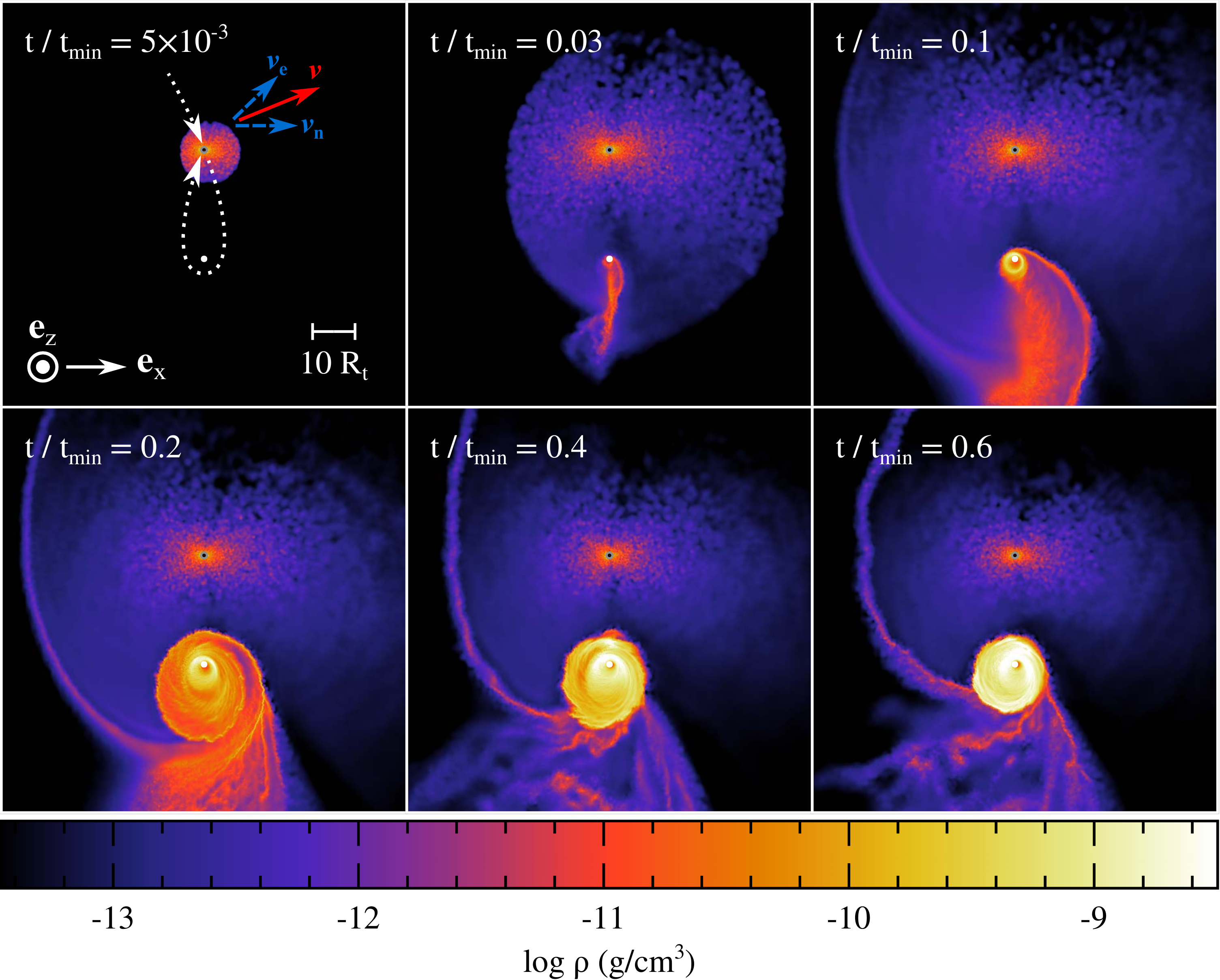}
\caption{Snapshots showing the gas density for simulation A in a slice parallel to the stellar orbital plane that contains the black hole at different times of $t/\tmin = 5 \times 10^{-3}$, 0.03, 0.1, 0.2, 0.4 and 0.6. The value of the density increases from black to white, as shown on the colour bar. The white segment on the first snapshot indicates the scale used and the orientation is given by the white arrows. The black hole is depicted by the white dot and the grey circle represents the intersection point, from which matter is injected into the computational domain. On the first snapshot, the white dotted line displays the precessed trajectory of the returning stream. The red solid arrow represents the velocity of a given outflowing particle while the blue dashed ones show its net and expansion components.}
\label{fig:density_plane_relativistic}
\end{figure*}

 The velocity field of the outflowing gas injected into the simulations is set to $\vect{v} = \vect{v}_{\rm e} + \vect{v}_{\rm n}$ that is the vectorial sum of the expansion and net speeds defined in Section \ref{sec:initial}. Because of adiabatic losses from the self-crossing shock, the radiation and thermal components of its internal energy are set to small values such that this expanding matter is highly supersonic. The positions of the injected particles in the sphere surrounding the intersection point are chosen in order to reproduce the density of the outflowing gas found by \citet{lu2020}. The azimuthal angle is obtained from a uniform random distribution to reflect the symmetry of the outflow along this direction. Instead, the polar dependence is specified by the probability distribution $\diff p / \diff \vartheta = 2 \pi \bar{\rho}_{\rm sc} (\vartheta) \sin \vartheta$, making use of equation \eqref{eq:density-profile}. Every time a particle is injected, its polar angle is set to $\vartheta = F^{-1} (s)$ where $s$ is a uniformly random number in the range $0\leq s \leq 1$ and the function $F^{-1}$ is the inverse of $s = F(\vartheta) = 2 \pi \int^{\vartheta}_0 \bar{\rho}_{\rm sc} (\vartheta') \sin \vartheta' \diff \vartheta'$ \citep{press1992} that we evaluate numerically. The resulting density of the injected gas can be seen at early times in the first snapshot of Fig. \ref{fig:density_plane_relativistic}, where the precessed trajectory of the returning stream is displayed with the white dotted line. The red solid arrow represents the velocity of a given outflowing particle while the blue dashed ones show its net and expansion components.

After the debris has been injected into the computational domain, we start to follow its evolution by solving the radiation-hydrodynamics equations given in the gas co-moving frame by \citep{mihalas1984}

\be
\rho  \frac{\diff \vect{v}}{\diff t} = -\nabla P_{\rm g} + \rho \kappa_{\rm s} \frac{\vect{F}_{\rm r}}{c} - \rho \nabla \Phi_{\rm h},
\label{eq:momentum}
\ee
\be
\rho \frac{\diff}{\diff t} \left( \frac{e_{\rm g}}{\rho} \right) = - \frac{P_{\rm g}}{\rho} \nabla \cdot \vect{v} - 4 \pi j_{\rm e} + \rho \kappa_{\rm a} c \, e_{\rm r} ,
\label{eq:gas_energy}
\ee
\be
\rho \frac{\diff}{\diff t} \left( \frac{e_{\rm r}}{\rho} \right) =  - \tens{P}_{\rm r}:\nabla \vect{v} - \nabla \cdot \vect{F}_{\rm r} + 4 \pi j_{\rm e} - \rho \kappa_{\rm a} c \, e_{\rm r},
\label{eq:radiation_energy}
\ee
\be
\frac{\rho}{c^2} \frac{\diff}{\diff t} \left( \frac{\vect{F}_{\rm r}}{\rho} \right) = - \nabla \cdot \tens{P}_{\rm r} - \rho \kappa_{\rm s} \frac{\vect{F}_{\rm r}}{c},
\label{eq:radiation_flux}
\ee
which are valid at first order in the ratio $v/c$ of gas velocity to speed of light. Here, $\rho$, $\vect{v}$, $e_{\rm g}$ and $P_{\rm g} = 2 e_{\rm g}/3$ represent the gas density, velocity, thermal energy density and pressure, respectively. The frequency-integrated radiation energy density, flux and pressure tensor are denoted by $e_{\rm r}$, $\vect{F}_{\rm r}$ and $\tens{P}_{\rm r}$. As we show below, scattering opacity dominates that due to absorption with $\kappa_{\rm s}\gg\kappa_{\rm a}$ such that we do not consider the influence of the latter on the process of photon diffusion. The gravitational force from the black hole is modelled using a potential $\Phi_{\rm h}$ while  $j_{\rm e}$ is the emission coefficient.

For an optically thick gas, the radiation field is close to isotropic that implies the relation $\tens{P}_{\rm r} = (e_{\rm r}/3) \tens{I}$. Using equation \eqref{eq:radiation_flux} in steady-state, the radiation flux can then be evaluated as $\vect{F}_{\rm r} = - c\nabla e_{\rm r}/(3 \rho \kappa_{\rm s})$ that leads to a diffusion equation for the radiation energy after combining with equation \eqref{eq:radiation_energy}. However, this expression allows for an arbitrarily fast transport of radiation if the gas is optically thin, which is unphysical. We therefore evaluate the flux instead as
\be
\vect{F}_{\rm r} = - \frac{\lambda c}{\rho \kappa_{\rm s}} \nabla e_{\rm r},
\label{eq:flux_limited}
\ee
which is known as the FLD approximation and replaces equation \eqref{eq:radiation_flux} in the above system. Following \citet{levermore1981}, the flux-limiter $\lambda \leq 1/3$ is specified by
\be
\lambda = \frac{2+R}{6+3R+R^2},
\label{eq:flux_limiter}
\ee
where $R = |\nabla e_{\rm r}|/(\kappa_{\rm s} \rho e_{\rm r})$. According to this expression, radiation propagates at the speed of light with $|\vect{F}_{\rm r}| = c e_{\rm r}$ in optically thin regions where $R \gg 1$ that corresponds to the free-streaming regime. For large optical depths where $R \ll 1$, the diffusion limit is also recovered since $\lambda \approx 1/3$. Finally, the radiation pressure tensor is determined from the relation
\be
\tens{P}_{\rm r} = \lambda e_{\rm r} \tens{I},
\label{eq:pressure_tensor}
\ee
which correctly captures the isotropy of the radiation field in optically thick regions. The FLD approximation corresponding to equation \eqref{eq:flux_limited} is known to result in unphysical behaviours in optically thin regions. An important source of error relates to the direction of propagation that is set by the gradient of radiation energy while photons should move in straight lines independent of the local gas properties. Similarly, equation \eqref{eq:pressure_tensor} implies that the pressure tensor is isotropic, which only applies to the diffusion regime.\footnote{We note that this formulation has been used in several existing radiation-hydrodynamics codes \citep{commercon2011,chatzopoulos2019} although other implementations \citep{turner2001,whitehouse2004} adopt expressions that are more accurate in the free-streaming limit.} Nevertheless, this approximate treatment is legitimate for our problem since the vast majority of the gas has very high optical depths, as we further demonstrate in Section \ref{sec:emerging}.

\begin{figure*}
\centering
\includegraphics[width=\textwidth]{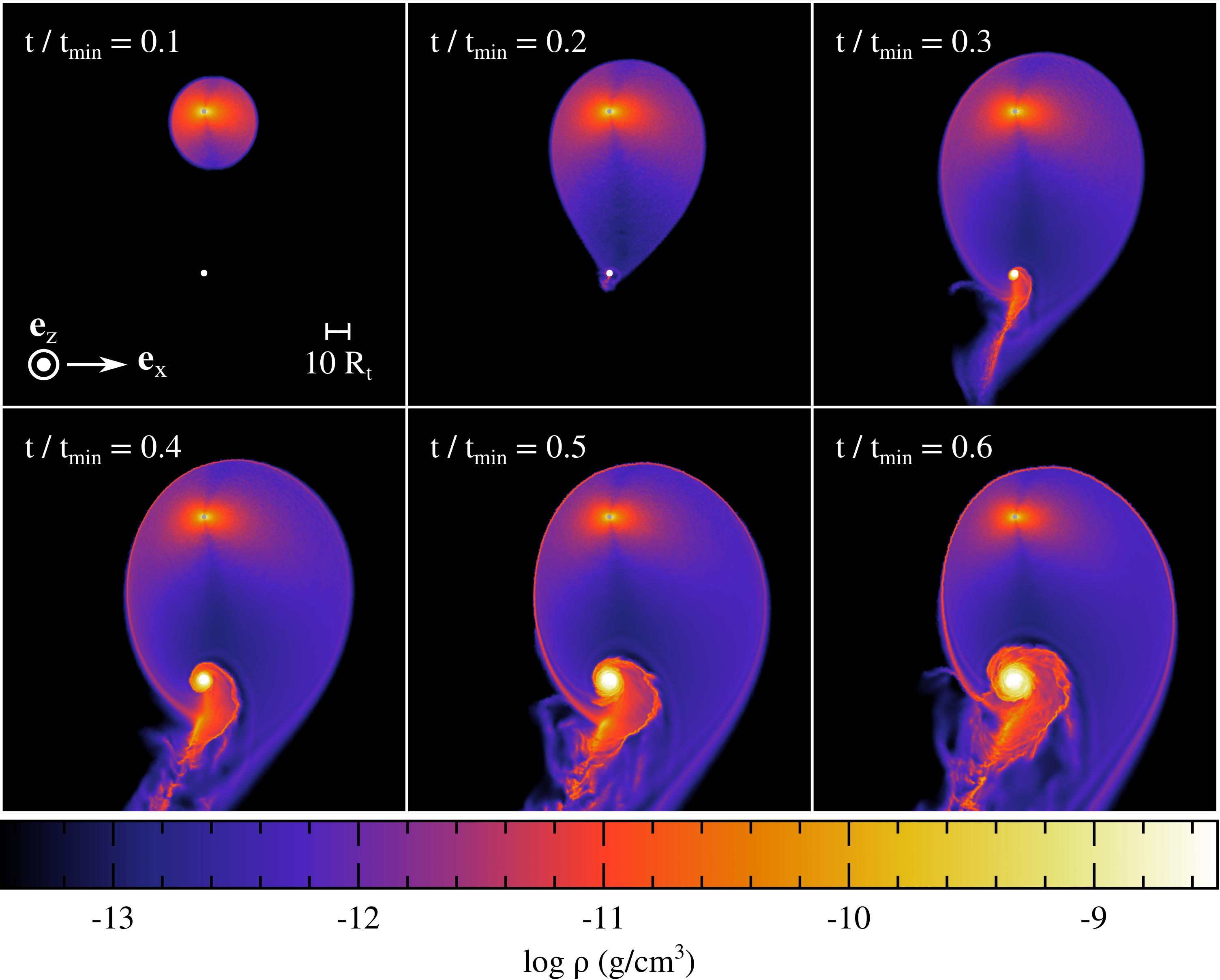}
\caption{Snapshots showing the gas density for simulation B in a slice parallel to the stellar orbital plane that contains the black hole at different times of $t/\tmin = 0.1$, 0.2, 0.3, 0.4, 0.5 and 0.6. The value of the density increases from black to white, as shown on the colour bar. The white segment on the first snapshot indicates the scale used and the orientation is given by the white arrows. The black hole is depicted by the white dot and the grey circle represents the intersection point, from which matter is injected into the computational domain.}
\label{fig:density_plane_non_relativistic}
\end{figure*}

\begin{figure*}
\centering
\includegraphics[width=\textwidth]{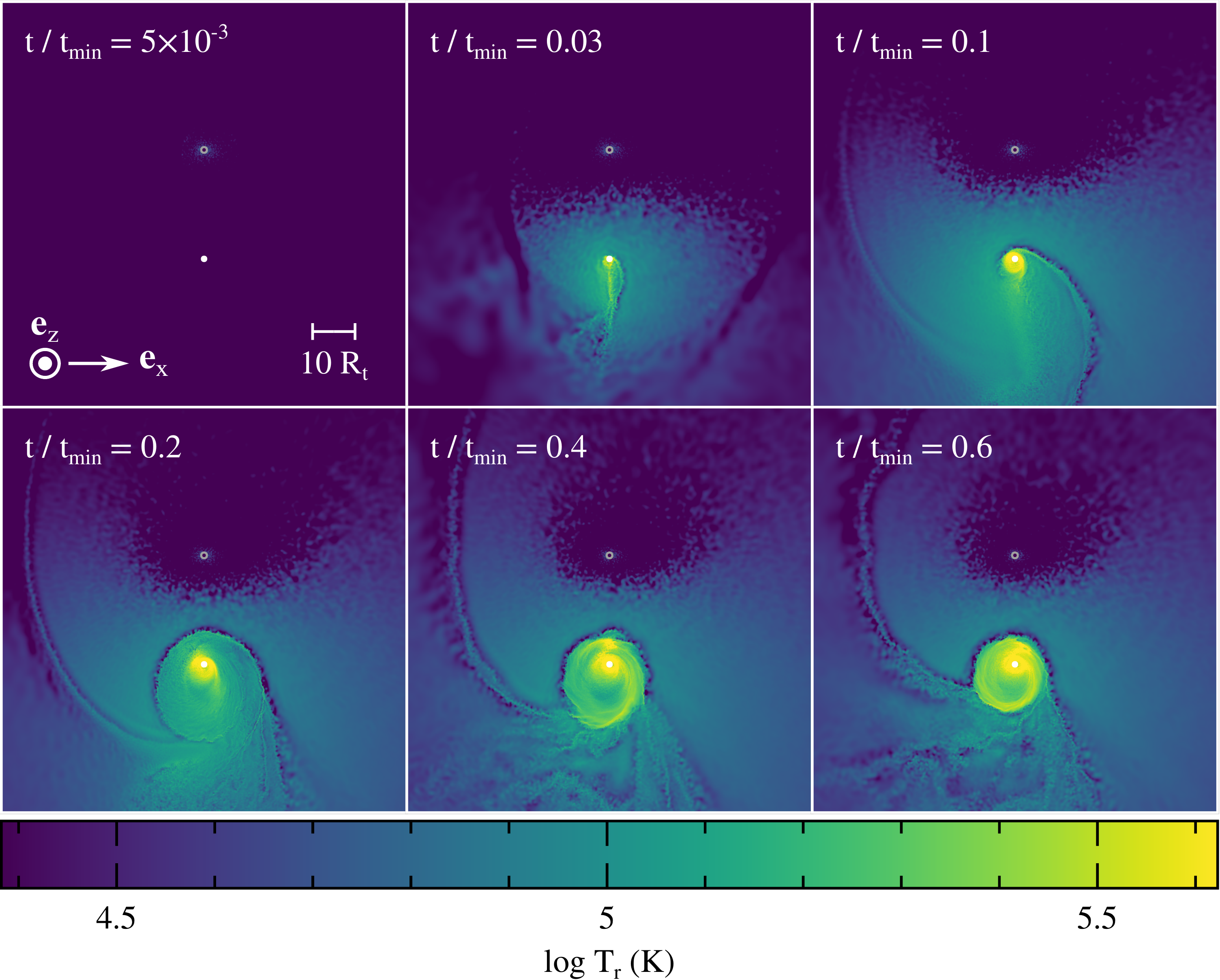}
\caption{Snapshots showing the radiation temperature $T_{\rm r}= (e_{\rm r}/a)^{1/4}$ for simulation A in a slice parallel to the stellar orbital plane that contains the black hole at different times of $t/\tmin = 5 \times 10^{-3}$, 0.03, 0.1, 0.2, 0.4 and 0.6. The value of the temperature increases from purple to yellow, as shown on the colour bar. The white segment on the first snapshot indicates the scale used and the orientation is given by the white arrows. The black hole is depicted by the white dot and the grey circle represents the intersection point, from which matter is injected into the computational domain.}
\label{fig:temperature_plane_relativistic}
\end{figure*}

\begin{figure*}
\centering
\includegraphics[width=\textwidth]{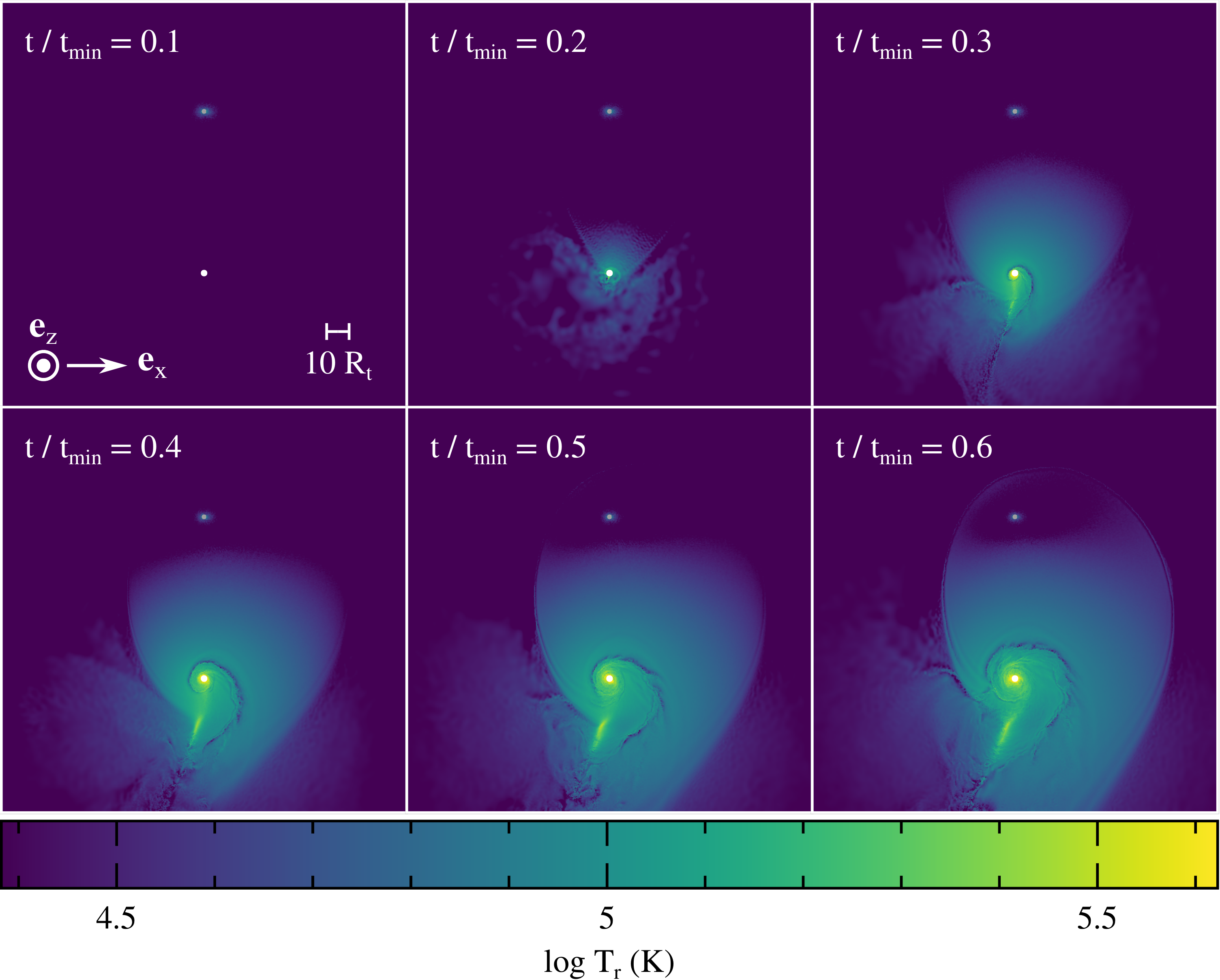}
\caption{Snapshots showing the radiation temperature $T_{\rm r}= (e_{\rm r}/a)^{1/4}$ for simulation B in a slice parallel to the stellar orbital plane that contains the black hole at different times of $t/\tmin = 0.1$, 0.2, 0.3, 0.4, 0.5 and 0.6. The value of the temperature increases from purple to yellow, as shown on the colour bar. The white segment on the first snapshot indicates the scale used and the orientation is given by the white arrows. The black hole is depicted by the white dot and the grey circle represents the intersection point, from which matter is injected into the computational domain.}
\label{fig:temperature_plane_non_relativistic}
\end{figure*}

We adopt the electron scattering opacity $\kappa_{\rm s} = 0.34 \, \cm2g$, which corresponds to a typical stellar hydrogen abundance. In optically thick regions, this interaction specifies the rate at which radiation moves with respect to the gas according to the diffusion equation. If the gas is surrounded by complete vacuum devoid of particles, we find that radiation accumulates at this boundary and may continue to influence the gas motion. This is unphysical since the photons should realistically leave the system entirely by propagating through some optically thin surrounding matter. Numerically, we prevent this artefact by filling the computational domain with a tenuous bath made of $N_{\rm b} = 10^5$ low-mass particles with a density profile of $\rho_{\rm b} (R) = \rho_{\rm b,0}$ for $R \leq R_{\rm br}$ and $\rho_{\rm b} (R) = \rho_{\rm b,0} (R/R_{\rm br})^{-3}$ for $R_{\rm br} \leq R \leq R_{\rm out}$, where $\rho_{\rm b,0} = 10^{-18} \gcm3$, $R_{\rm br} = R_{\rm t}$ and $R_{\rm out} = 1 \, \rm ly$. This value of the outer radius ensures that free-streaming radiation escaping the system at early times is still propagating outward at the end of the simulation when $t = 0.5 \, \tmin \ll 1 \, \rm yr$ without reaching the edge of the bath. The slope at $R \geq R_{\rm br}$ is chosen because it leads to a fixed mass per logarithmic bins of radius such that bath particles are present at large radii despite the very low density there, as necessary to keep transporting the radiation energy. The total mass of the bath is $\sim 4 \pi \rho_{\rm b,0} R_{\rm br}^3 \ln (R_{\rm out}/R_{\rm br}) \approx 10^{-8} \msun$ that is much lower than that of the stellar matter around the black hole at all times, and has therefore no impact on its dynamical evolution. To prevent the bath particles from collapsing to small radii over the course of the simulation, we make them insensitive to the gravitational attraction of the black hole.

We treat the combined impact of thermal and radiation pressure through a total pressure given by $P_{\rm t} = P_{\rm g} + \lambda e_{\rm r} /3$. This is possible since equations \eqref{eq:pressure_tensor} and \eqref{eq:radiation_flux} in steady-state can be combined to rewrite the acceleration due to radiation as $\rho \kappa_{\rm s} \vect{F}_{\rm r}/c \approx  \nabla (\lambda e_{\rm r} /3)$ in equation \eqref{eq:momentum}. 
Similarly, the relation $\tens{P}_{\rm r} : \nabla \vect{v} \approx \lambda e_{\rm r} \nabla \cdot \vect{v}/3$ can be used to combine the terms due to pressure in equations \eqref{eq:gas_energy} and \eqref{eq:radiation_energy} into a single one that determines the evolution of the total energy density $e_{\rm t} = e_{\rm g} + e_{\rm r}$. The code updates the gas velocity and total energy due to the combined influence of thermal and radiation pressure. Non-adiabatic heating is attributed to shocks that only increases the thermal energy while the radiation energy density is recovered from $\diff (e_{\rm r}/\rho) / \diff t = - \lambda e_{\rm r} \nabla \cdot v / 3$ that takes into account its adiabatic variation.\footnote{In the diffusive regime where $\lambda \approx 1$, combining this equation with that of mass conservation $\diff \rho / \diff t = -\rho \nabla \cdot \vect{v}$ leads to $\diff (e_{\rm r}/\rho) / \diff t = e_{\rm r}/(3 \rho) \diff \rho / \diff t$. This yields $e_{\rm r} \propto \rho^{4/3}$ that is the correct scaling in the co-moving frame when radiation is adiabatically evolved through advection with the gas in the absence of radiative diffusion.} We emphasize that \textsc{gizmo} \citep{hopkins2015} naturally treats shocks by solving the Riemann problem at the faces reconstructed around each particle. This technique is different from that of smoothed-particle-hydrodynamics adopted in our earlier work \citep{bonnerot2020-realistic} where dissipation was taken into account through artificial viscosity. Nevertheless, we find that the gas evolution is essentially identical using the two numerical methods when gas adiabaticity is assumed. This preliminary check allows us to confidently associate the new features found in the simulations presented here with the additional impact of radiative processes.

\begin{figure*}
\centering
\includegraphics[width=\columnwidth]{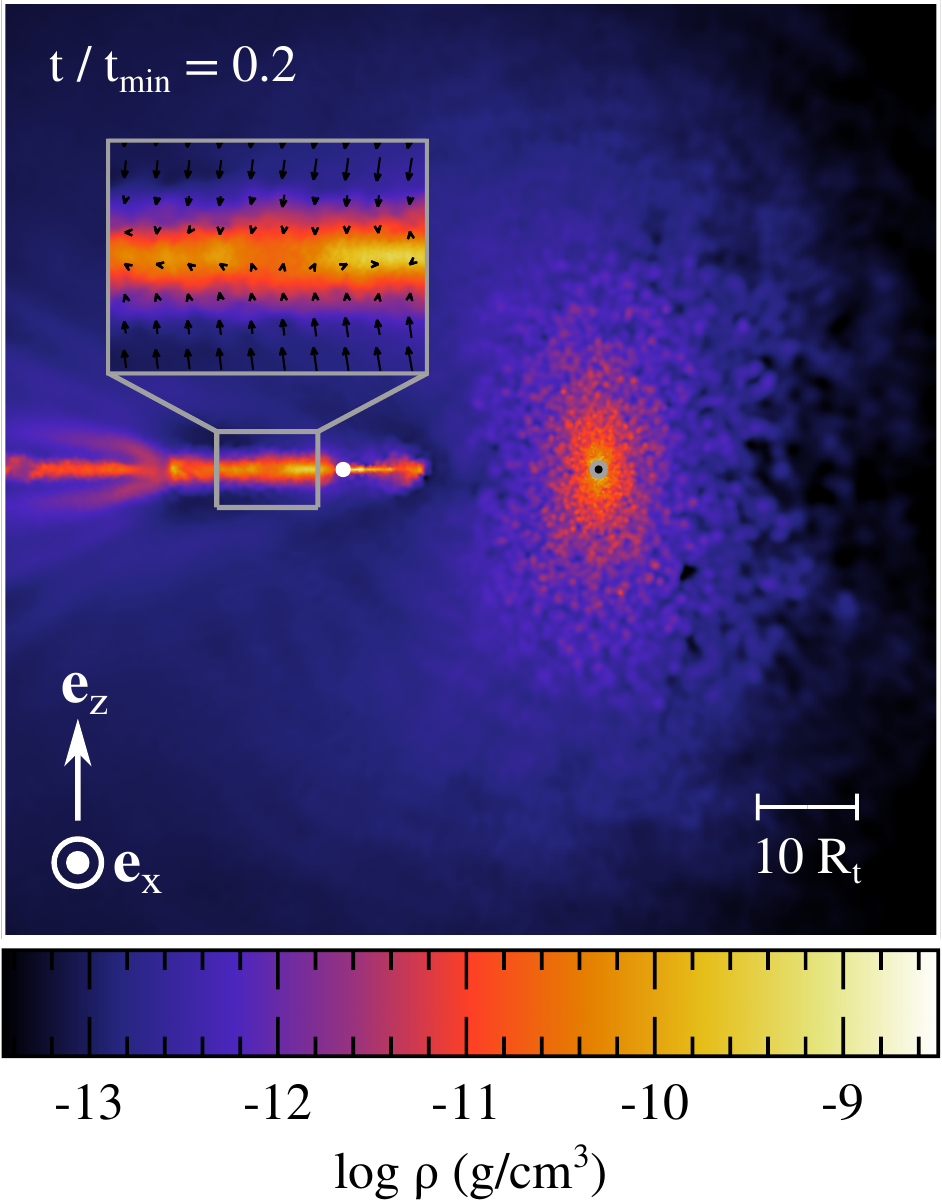}
\hspace{0.5cm}
\includegraphics[width=\columnwidth]{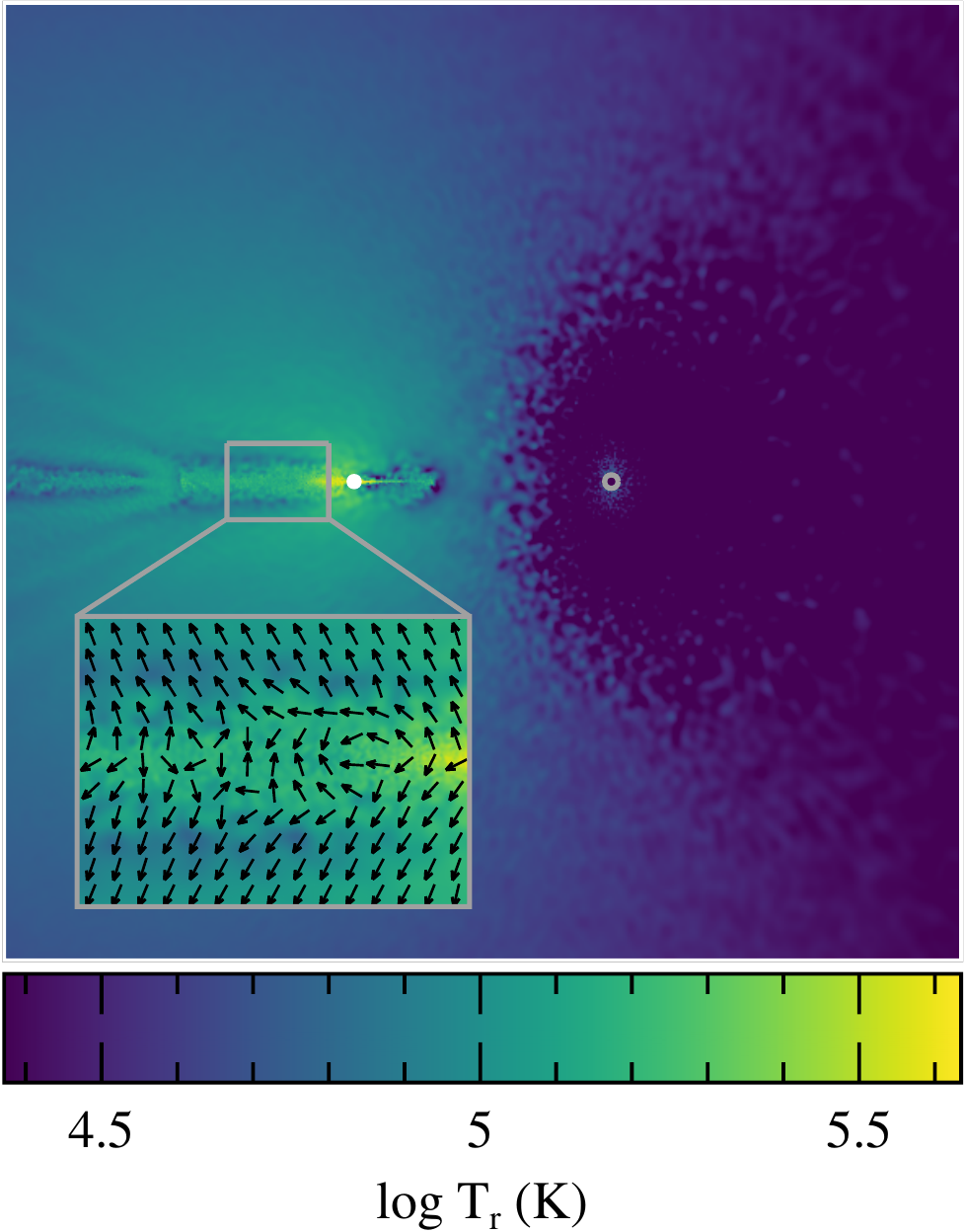}
\caption{Gas density (left panel) and radiation temperature (right panel) in a slice orthogonal to the stellar orbital plane containing the black hole for simulation A at $t/\tmin = 0.2$. The density increases from black to white and the radiation temperature from purple to yellow, as shown on the colour bars. The white segment indicates the scale used and the orientation is given by the white arrows. The black hole is depicted by the white dot and the grey circle represents the intersection point, from which matter is injected into the computational domain. The two insets show zoom-ins on the same region of high density near the mid-plane corresponding to the grey rectangles. The black arrows represent the velocity field and the direction of the diffusive flux on the left and right panel, respectively.}
\label{fig:density_temperature_vertical_relativistic}
\end{figure*}

The last two terms in equations \eqref{eq:gas_energy} and \eqref{eq:radiation_energy} correspond to emission and absorption that are dominated by free-free interactions in the physical conditions considered.\footnote{It is possible that bound-free opacity becomes comparable to that provided by free-free interactions, especially in regions of low temperatures where more of the gas can become neutral. Once the radiation energy density becomes sufficiently large, the inverse-Compton process may also participate to the energy exchange between matter and radiation. It could also be accompanied by synchrotron radiation if the magnetic field becomes strong enough, due to dynamo processes taking place near the black hole. However, we show below that free-free processes by themselves are already able to efficiently drive gas and radiation towards temperature equilibrium. Indeed, equation \eqref{eq:equilibrium} below shows that the associated timescale is shorter than all the other relevant ones for most of the gas densities and temperatures found in our simulations. Even though including bound-free opacity as an additional source of absorption may further accelerate this process, this modification is therefore unlikely to change the properties of the radiation.} We therefore use the emission coefficient for this process given by $j_{\rm ff} = 20 \, \ergcm3s (T_{\rm g} / 10^5 \kelvin )^{1/2} (\rho/10^{-11} \gcm3)^{2}$ \citetext{equation (10.12) of \citealt{draine2011}}, where the gas temperature is given by the ideal gas relation $P_{\rm g} = \rho k_{\rm B} T_{\rm g}/(m_{\rm p} \mu)$. Here, $\mu = 0.61$ denotes the molecular weight for a fully-ionized gas and $m_{\rm p}$ is the proton mass. Under local thermodynamical equilibrium, the Planck mean of the absorption opacity is obtained by integrating Kirchhoff's law as $\kappa_{\rm a} = j_{\rm ff} /(\rho B)$ where $B = a c T^4_{\rm g}/(4 \pi)$ denotes the integrated Planck function. Numerically, this yields
\be
\kappa_{\rm a} = 9 \times 10^{-4} \cm2g  \left(\frac{\rho}{10^{-11} \gcm3}\right) \left( \frac{T_{\rm g}}{10^5 \kelvin} \right)^{-7/2},
\ee
using values of the gas density and temperature typical of our problem. Absorption opacity is therefore negligible compared to electron-scattering opacity with $\kappa_{\rm a}\ll\kappa_{\rm s}$ that justifies to neglect its influence on photon diffusion, as assumed in the system of equations solved by the code. As we show below, absorption and emission typically take place on a timescale much shorter than that $\Delta t$ given by all the other processes we consider. In order to avoid significant computational overhead, these effects are treated independently that amounts to computing the variation of the radiation and thermal energy densities over a timestep $\Delta t$ according to 
\be
\left(\frac{\diff e_{\rm r}}{\diff t} \right)_{\rm ab}  = a c  \rho \kappa_{\rm a} (T^4_{\rm g} - T^4_{\rm r}),
\label{eq:absorption}
\ee
and energy conservation
\be
e_{\rm g} + e_{\rm r} = e^0_{\rm t}.
\label{eq:conservation}
\ee
where this first equation corresponds to the last two terms of equation \eqref{eq:radiation_energy} associated with absorption and emission. Here, the radiation temperature is defined by $T_{\rm r} = (e_{\rm r}/a)^{1/4}$ while $e^0_{\rm t}$ denotes the total energy density at the beginning of the timestep. Equation \eqref{eq:absorption} corresponds to an evolution towards temperature equilibrium with $T_{\rm g} = T_{\rm r}$ that is reached on a timescale of about $t_{\rm eq} = \max(e_{\rm g},e_{\rm r})/|a c  \rho \kappa_{\rm a} (T^4_{\rm g} - T^4_{\rm r})|$. In the situation of interest where hot gas generates radiation, it can be numerically evaluated as
\be
t_{\rm eq} \approx  \frac{e_{\rm g}}{a c \rho \kappa_{\rm a} T^4_{\rm g}} = 300 \, {\rm s}  \left(\frac{\rho}{10^{-11} \gcm3}\right)^{-1} \left( \frac{T_{\rm g}}{10^{10} \kelvin} \right)^{1/2},
\label{eq:equilibrium}
\ee
for typical values of the gas temperature and density immediately after the shocks captured in our simulation, which confirms that it is shorter than other physical processes. For instance, the dynamical time only decreases to a about a thousand seconds at the tidal radius that is close to the inner edge of our computational domain. For this reason, we make use of the following numerical approach to treat this effect. If $\Delta t / t_{\rm eq} > 100 \gg 1$ as should be the case for most of the gas, we consider that equilibrium is reached at the end of the timestep. This allows us to find the updated energies from the equilibrium temperature $T_{\rm eq}$ that we can rapidly obtain by analytically solving the quartic equation $a T^4_{\rm eq} + 3 \rho k_{\rm B} T_{\rm eq} /(2 \mu) = e^0_{\rm t}$. If instead $\Delta t < 100 \, t_{\rm eq}$, we numerically solve equations \eqref{eq:absorption} and \eqref{eq:conservation} to update the energies. This sub-timestepping routine allows us to treat the evolution towards temperature equilibrium through free-free interactions self-consistently without drastically shortening the timesteps that would otherwise slow down the simulation.

The gravity of the black hole is assumed to be Keplerian by setting the potential in equation \eqref{eq:momentum} to $\Phi_{\rm h} = -G \mh/R$. This choice is justified by the negligible differences seen in our previous work \citep{bonnerot2020-realistic} between simulations including  general-relativistic effects and using a Keplerian treatment. The reason is that most collisions experienced by the gas result from the wide range of trajectories it follows after passing through the self-crossing shock. These interactions are largely independent on the relatively small impact of general relativity on the orbits of the stellar debris. Accretion is taken into account by removing from the simulation the particles that reach a distance to the black hole less than $R_{\rm acc} = 6 \rg$, which is the inner-most stable circular orbit.

\section{Results}

\label{sec:results}

We now describe the combined evolution of gas and radiation following the self-crossing shock as the debris keeps interacting to form an accretion disc around the black hole.\footnote{Movies of the simulations are available online at \url{http://www.tapir.caltech.edu/~bonnerot/first-light.html}.} Simulation A is often used to describe the impact of radiative processes since its results can be directly compared with our earlier investigation \citep{bonnerot2020-realistic} that uses the same parameters but assumed gas adiabaticity. While describing the gas evolution found in simulation B, we emphasize the most important differences induced by the less relativistic encounter it considers.

\begin{figure}
\centering
\includegraphics[width=\columnwidth]{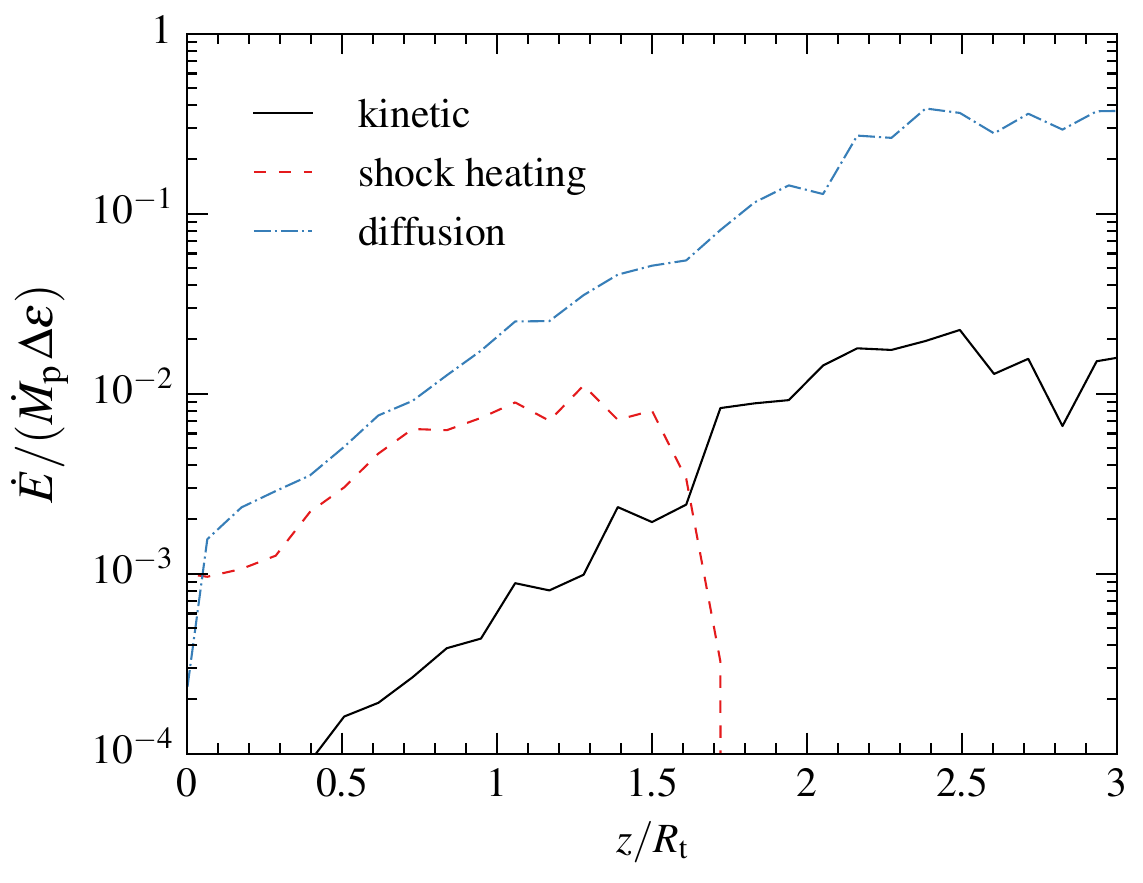}
\caption{Vertical profile of different rates of energy variation evaluated as the disc forms at $t = 0.2\, \tmin$ by considering the gas contained between the radius $R_{\rm in} = 8\rt$ and $R_{\rm out} = 12 \rt$ within an azimuthal angle $\Delta \varphi = \pi/8$ from the $-\ey$ direction. The black solid line denotes the rate of inward kinetic energy transport due to the infalling matter computed as $\dot{E}_{\rm kin} = \dot{M}_{\rm in} v^2_{\rm z}/2$, where the vertical component $v_{\rm z}$ of the gas velocity and the vertical net inflow rate $\dot{M}_{\rm in}$ are obtained from the simulation. The red dashed line has its shape given by the thermal energy increase due to shocks in each vertical layers, not taking into account cooling by free-free emission. These contributions are added up for vertical distances in the range $0\leq z \leq 3 \rt$ to obtain the total heating rate, which is used to fix the peak of this curve. The dot-dashed blue line corresponds to the rate of outward diffusion obtained from $\dot{E}_{\rm dif} =  (\vect{F}_{\rm r} \cdot \ez) \, \Delta A$ by multiplying the vertical component of the radiation flux by the surface $\Delta A = (R^2_{\rm out}-R^2_{\rm in}) \Delta \varphi$.}
\label{fig:edotvsz}
\end{figure}

\begin{figure}
\centering
\includegraphics[width=\columnwidth]{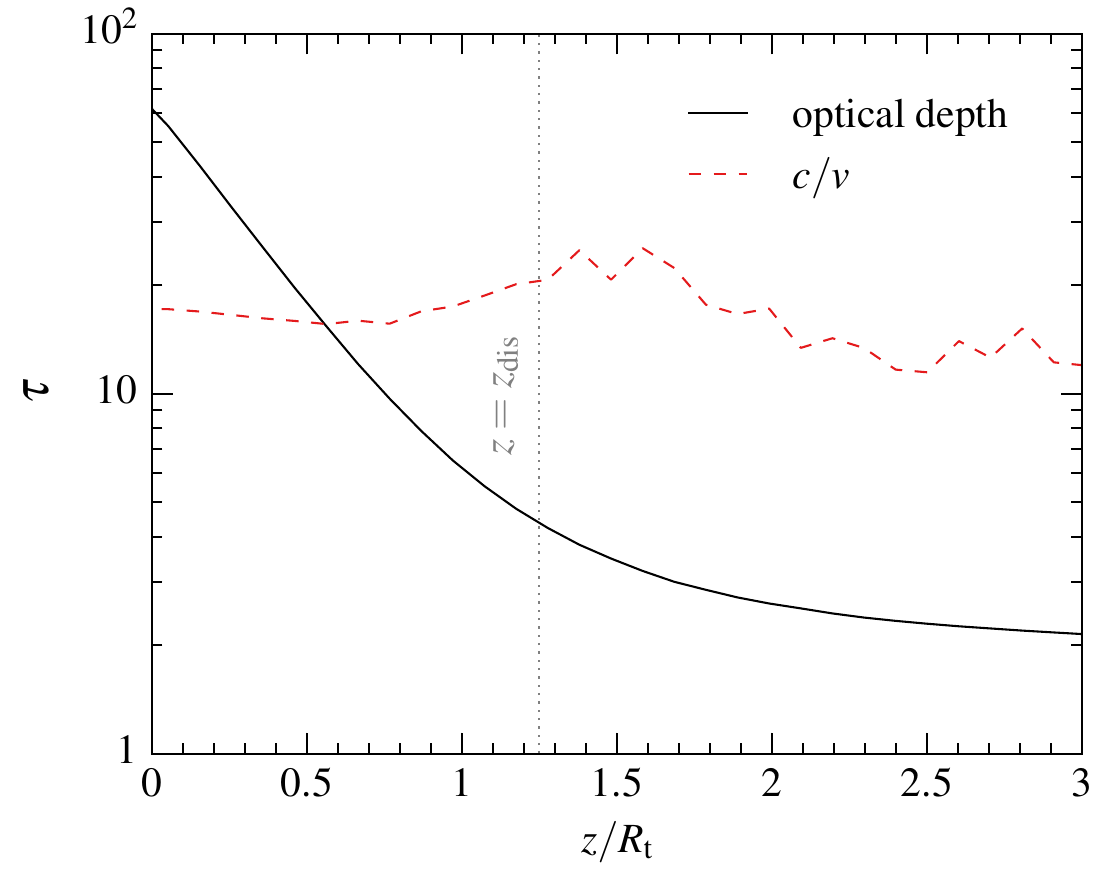}
\caption{Vertical profile of the integrated optical depth (black solid line) obtained at $t = 0.2\, \tmin$ from $\tau = \int_{z}^{z_{\rm max}} \rho \kappa_{\rm s} \diff z'$, setting the upper bound to $z_{\rm max} = 10\rt$. It is calculated by considering the gas contained between the radius $R_{\rm in} = 8\rt$ and $R_{\rm out} = 12 \rt$ within an azimuthal angle $\Delta \varphi = \pi/8$ from the $-\ey$ direction. The grey dotted vertical line denotes the height $z_{\rm dis} = 1.25 \rt$ corresponding to the surface of the accumulated gas where heating by shocks is the largest, as can be seen from Fig. \ref{fig:edotvsz} that focuses on the same region. The red dashed line corresponds to the ratio $c/v$ of speed of light to gas velocity that we evaluate from the simulation.}
\label{fig:tauvsz}
\end{figure}

\subsection{Radiative cooling at secondary shocks}

\label{sec:secondary}

Following its injection inside the computational domain, the outflowing debris expands in a quasi-spherical manner, as specified by the properties of the self-crossing shock. The evolution of this gas can be followed from the snapshots of Figs. \ref{fig:density_plane_relativistic} and \ref{fig:density_plane_non_relativistic} that show its density in a slice parallel to the stellar orbital plane containing the black hole for simulations A and B, respectively. Owing to its large range of trajectories, the expanding envelope intersects itself near the mid-plane on the side of the black hole opposite to the injection point. The first interactions involve the gas injected with inward and nearly radial motion that is the first to reach this location. Due to the larger intersection radius of the less relativistic encounter considered in simulation B, this initial collision takes place later than in simulation A since additional time is required for the injected matter to reach small distances. As gas keeps infalling near the black hole, more of these secondary shocks take place, whose impact on the hydrodynamics is studied below. Most importantly, these interactions cause a circularization of the trajectories into a nascent accretion disc due to the associated kinetic energy dissipation, a process examined in detail in Section \ref{sec:circularization}. 

The shocked debris experiences an increase of its temperature to $T_{\rm g} \approx m_{\rm p} v^2_{\rm sh} / k_{\rm B} = 10^{10} \kelvin$ where $v_{\rm sh} \approx 0.03 c$ is the typical relative speed involved. A fast rise of the radiation temperature $T_{\rm r} = (e_{\rm r} / a)^{1/4}$ ensues due to free-free emission, which is displayed in the snapshots of Figs. \ref{fig:temperature_plane_relativistic} and \ref{fig:temperature_plane_non_relativistic} that focus on the same parallel slice as before for the two simulations. As evaluated in equation \eqref{eq:equilibrium}, equilibrium with matter is reached on a very short timescale of $t_{\rm eq} \lesssim 10^{-3} \tmin$ that results in a radiation temperature of order $T_{\rm r} \approx (\rho v^2_{\rm sh}/a )^{1/4}=10^5 \kelvin$ for a typical density of $\rho \approx 10^{-11} \gcm3$. Throughout the colliding gas, pressure is dominated by radiation with a ratio to the thermal component of $P_{\rm r}/ P_{\rm g} \gtrsim  1000$. It can be seen that the radiation temperature also increases away from the dissipation sites, which is due to radiative diffusion that progressively irradiates the previously cold surrounding matter. As we describe more precisely in Section \ref{sec:emerging}, this process continues until photons are transported out to large distances through the injected debris. Part of this radiation eventually emerges from the system where it may keep propagating to an external observer.

Several properties of the shocked matter indicate that radiative diffusion has a significant impact on the dynamical evolution due to the local energy depletion it induces. When the first debris arrives near the black hole, their intersection results in the formation of a narrow filament of high density visible at $t/\tmin \approx 0.03$ in Fig. \ref{fig:density_plane_relativistic}, which corresponds to the interface between fluid elements arriving in opposite directions where their angular momenta partially cancel. Such a region of enhanced density is not expected from adiabatic shocks, which suggests that the gas evolution is closer to an isothermal one. This deviation from adiabaticity is even more evident from the vertical gas distribution displayed for simulation A in the left panel of Fig. \ref{fig:density_temperature_vertical_relativistic} that shows its density at $t/\tmin = 0.2$ in a slice orthogonal to the stellar orbital plane containing the black hole. It can be seen that the debris falling along the $\pm \ez$ directions piles up around the mid-plane after undergoing shocks, which results in the formation of a thin structure elongated towards the $-\ey$ direction. The same evolution takes place in simulation B when the injected gas reaches the equatorial plane. It is in stark contrast with the results of our earlier adiabatic simulation \citep{bonnerot2020-realistic} where large pressure gradients caused the gas to quickly expand, leading to a more puffy distribution with lower densities. This already demonstrates that radiative diffusion is fast enough for the associated energy transport to significantly affect the hydrodynamics of the shocked gas.

To examine more in detail the influence of radiative cooling, we focus on the region near the mid-plane where infalling matter continuously joins the gas that has already arrived. As can be seen from the velocity field in the inset of Fig. \ref{fig:density_temperature_vertical_relativistic} (left panel), the debris decelerates when reaching this denser area, which is accompanied by a shock. Fig. \ref{fig:edotvsz} presents the associated energy transfer above the mid-plane at $z>0$ for the debris contained within an azimuthal angle $\Delta \varphi = \pi/8$ of the $-\ey$ direction and between radii $R_{\rm in} = 8 \rt$ and $R_{\rm out} = 12 \rt$ when disc formation is ongoing at $t/\tmin = 0.2$. The rate at which kinetic energy is transported inward (black solid line) is evaluated as $\dot{E}_{\rm kin} = \dot{M}_{\rm in} v^2_{\rm z}/2$, directly taking the vertical component $v_{\rm z}$ of the gas velocity from the simulation and computing the net vertical inflow rate $\dot{M}_{\rm in}$ from the motion of gas particles by using the same technique as in Section \ref{sec:circularization}. As it approaches from large heights, the gas kinetic energy remains constant until the surface of the accumulated debris, where it starts rapidly decreasing as matter loses its vertical motion. This loss of kinetic energy is associated with a sharp increase of the heating rate produced by shocks. It is visible from the red dashed line whose shape is given by the thermal energy increase due to shocks in the different layers, not taking into account cooling through free-free emission. These contributions are then added up to obtain the total shock heating rate used to fix the peak value of this curve, which is as expected similar to the deposited rate of kinetic energy. Most of this dissipation occurs at the surface of the denser gas distribution around $z \approx z_{\rm dis} = 1.25 \rt$ while the debris closer to the mid-plane undergoes a much lower level of heating.

As described above, this thermal energy is then quickly converted into radiation that proceeds to diffuse away. The corresponding radiation temperature is shown in the right panel of Fig. \ref{fig:density_temperature_vertical_relativistic} that focuses on the same orthogonal slice as the left one. In the inset, the black arrows indicate the direction of the local radiation flux that points away from the mid-plane outside the dissipative layer.  We compute the rate of energy diffusion from $\dot{E}_{\rm dif} =  (\vect{F}_{\rm r} \cdot \ez) \, \Delta A$ by multiplying the vertical component of the radiation flux of equation \eqref{eq:flux_limited} by the surface $\Delta A = (R^2_{\rm out}-R^2_{\rm in}) \Delta \varphi$ of the region considered in Fig. \ref{fig:edotvsz}. It is shown with the blue dotted line that increases away from the mid-plane as more radiation gets produced. Near the surface around $z_{\rm dis} = 1.25 \rt$, this rate of diffusion is similar to that of kinetic energy dissipation corresponding to the red dashed line, implying that the gas cools efficiently. The forming disc therefore acts as an almost perfectly reflective surface where most of the energy brought in by the infalling debris is lost in the form of radiation.\footnote{Above the disc, we note that the diffusive luminosity is significantly larger than the other two rates of Fig. \ref{fig:edotvsz}, which we attribute to radiation diffusing from lower radii inside the selected wedge of gas and additional energy generated by interactions outside this region that is brought through advection.}

\begin{figure}
\centering
\includegraphics[width=\columnwidth]{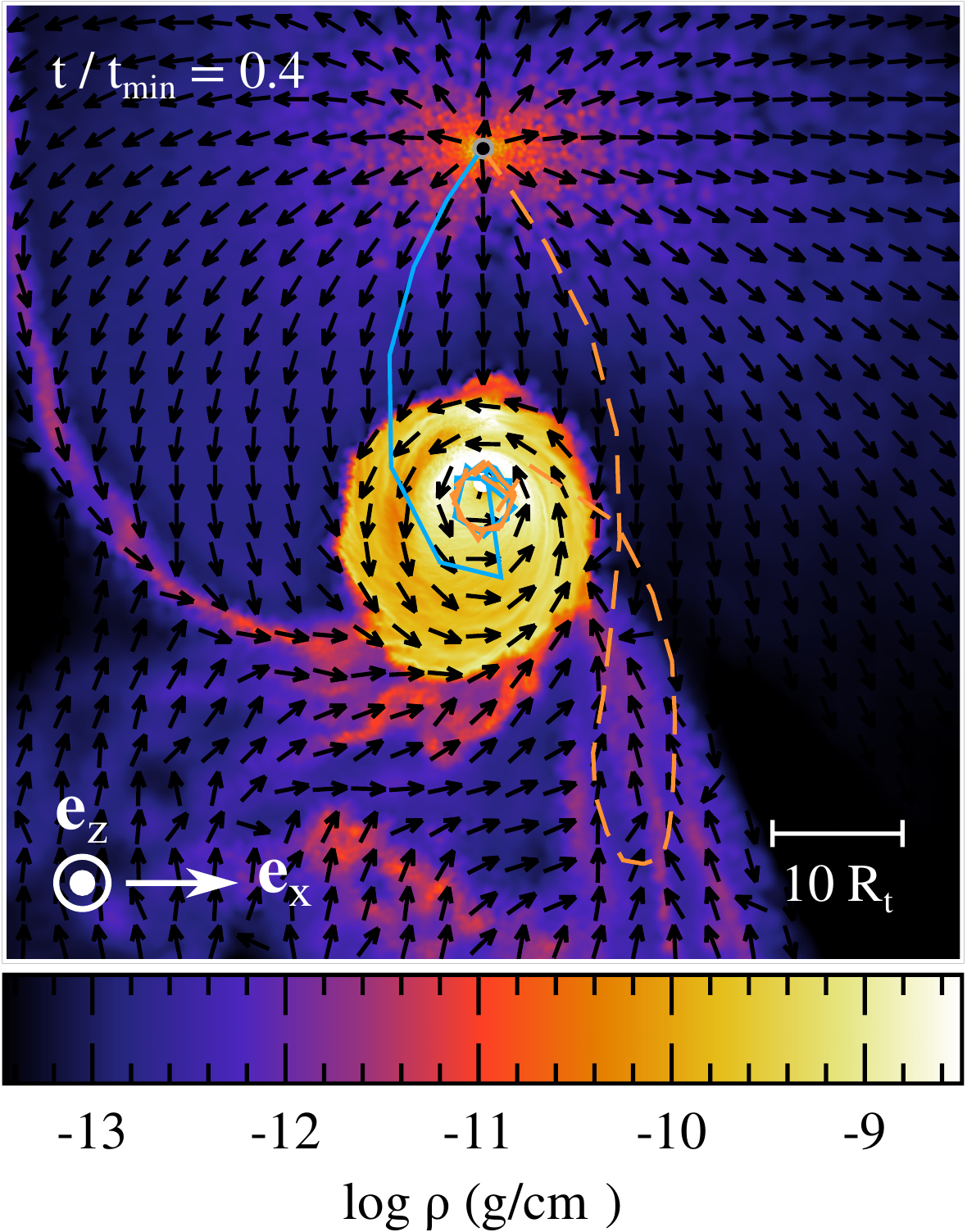}
\caption{Gas density in a slice parallel to the stellar orbital plane that contains the black hole at $t/\tmin = 0.4$ for simulation A. The value of the density increases from black to white, as shown on the colour bar. The white segment indicates the scale used and the orientation is given by the white arrows. The black hole is depicted by the white dot and the grey circle represents the intersection point, from which matter is injected into the computational domain. The black arrows of equal length display the direction of the velocity field. The two curves represent the trajectories of two particles that end up getting accreted after joining the disc directly from the intersection point (solid blue line) and following an interaction near the mid-plane at larger radii (dashed orange line).}
\label{fig:velocity_plane_relativistic}
\end{figure}

\begin{figure}
\centering
\includegraphics[width=\columnwidth]{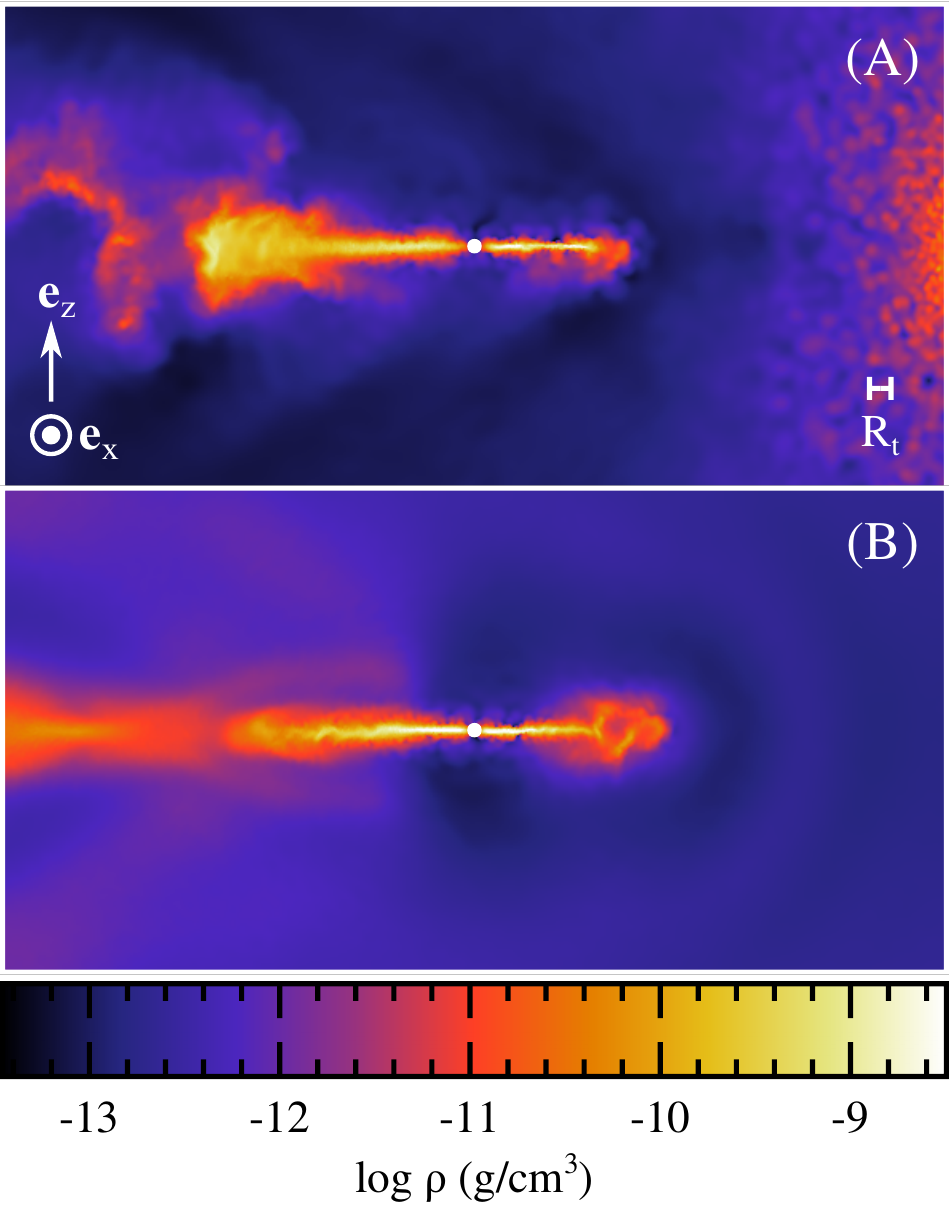}
\caption{Gas density in a slice perpendicular to the stellar orbital plane that contains the black hole when the discs have formed at $t/\tmin = 0.4$ for simulation A (upper panel) and $t/\tmin = 0.5$ for simulation B (lower panel). The value of the density increases from black to white, as shown on the colour bar. The black hole is depicted by the white dot. The white segments indicate the scale used and the orientation is given by the white arrows.}
\label{fig:density_vertical_comparison}
\end{figure}

The impact of this efficient cooling on the dynamics of the debris can be evaluated by comparing the diffusion and dynamical timescales whose ratio can be estimated as $t_{\rm dif} / t_{\rm dyn} \approx \tau v/c$. The optical depth is that encountered by the photons when diffusing away from the gas accumulated at the mid-plane, which we estimate by the integral $\tau = \int_{z}^{z_{\rm max}} \rho \kappa_{\rm s} \diff z'$, setting the upper bound to $z_{\rm max} = 10 \rt$. It is shown in Fig. \ref{fig:tauvsz} with the black solid line as a function of vertical distance for the same region as considered in Fig. \ref{fig:edotvsz} and compared to the ratio of speed of light to gas velocity $c/v$ (red dashed line). The fact that $t_{\rm dif} / t_{\rm dyn} \approx 0.2 \ll 1$ at the location of the dissipative layer around $z \approx z_{\rm dis}$ (grey dotted line) confirms that most of the heat generated there gets efficiently carried away by diffusion before it is able to affect the gas motion through radiation pressure. Note that considering only a fraction of the total velocity to estimate the above dynamical timescale strengthens this conclusion. The gas infalling near the black hole therefore experiences less expansion after undergoing shocks than in the adiabatic case, as we find in both simulations A and B. We show below that the properties of the disc forming at later times are affected in a similar way by this efficient radiative cooling.

\subsection{Circularization into a disc}

\label{sec:circularization}

As we just described, secondary shocks result in the formation of a thin gas distribution extending towards the $-\ey$ direction. Matter reaching this structure at large radii keeps  moving further away that contributes to the outflow discussed in Section \ref{sec:outflow}. Closer in, the shocked debris instead proceeds to approach the black hole where more interactions subsequently drive this gas towards circularization. The ensuing formation of an accretion flow can be followed from Figs \ref{fig:density_plane_relativistic} and \ref{fig:density_plane_non_relativistic}, which demonstrate that this process starts shortly after the first injected debris reaches the black hole vicinity. While being initially confined to small radii, the circularizing matter is joined by larger angular momentum gas that increases its radial extent until the properties of the newly-formed disc settle with an outer radius $R_{\rm d} \approx 10 \rt$ for both simulations A and B. The rapid disc formation found for the mildly relativistic encounter considered by simulation B is at odds with the common expectation that this process is slowed due to a weaker self-crossing shock. This is a consequence of the wide range of trajectories followed by the injected debris that promote strong secondary shocks despite the reduced dissipation occurring during the initial stream intersection. However, we emphasize that this conclusion as well as several properties of the formed disc sensitively depends on our simplified treatment of the self-crossing shock, whose validity is discussed in Section \ref{sec:self}.

In both simulations, we also find that the resulting disc rotates in the direction opposite to that of the star around the black hole. The corresponding velocity field is displayed for simulation A with black arrows in Fig. \ref{fig:velocity_plane_relativistic} that shows the density of the circularized debris at $t/\tmin = 0.4$. As explained in our previous study \citep{bonnerot2020-realistic}, this is naturally expected from the parameters used in this simulation based on the average angular momentum of the bound debris whose sign is opposite to that of the star due to the presence of unbound matter. However, this explanation cannot be directly applied to simulation B, for which all the injected matter is bound to the black hole. In this case, Fig. \ref{fig:density_plane_non_relativistic} shows that the fastest gas injected near the $\ex$ direction from the intersection point has its trajectory deflected by gravity towards lower radii. Nevertheless, these fluid elements remain far from the black hole such that its first interaction occurs at $R\approx 100 \rt$. Despite its co-rotation with the star, this gas therefore does not affect the angular momentum sign of the disc forming at shorter distances, which thus retains the counter-rotating motion of the more bound injected matter. Although this is not seen in our simulation, it is possible that the disc flips at later times as this matter eventually approaches the black hole. As we describe below, this gas also gets preferentially unbound by radiation pressure at large distances, which instead favours a situation where the disc retains the same angular momentum sign.

While the debris distribution rapidly puffed up in our previous adiabatic simulation \citep{bonnerot2020-realistic}, we find here that the nascent accretion disc remains much thinner. This is visible in Fig. \ref{fig:density_vertical_comparison} from the edge-on views of the formed disc shown at $t/\tmin = 0.4$ for simulation A (upper panel) and at $t/\tmin = 0.5$ for simulation B (lower panel). As explained in Section \ref{sec:secondary}, this reduced gas expansion results from efficient radiative cooling at the secondary shocks that prevents the internal energy from quickly building up inside the circularizing debris. An estimate of the width can be obtained from hydrostatic equilibrium that imposes an aspect ratio $H/R \approx c_{\rm s} / v_{\rm K}$ \citep{frank2002} where $c_{\rm s} = (a T^4_{\rm r}/3 \rho)^{1/2}$ and $v_{\rm K} = (G \mh /R)^{1/2}$ are the sound speed for a gas dominated by radiation pressure and the Keplerian velocity, respectively. For simulation A at $t/\tmin = 0.4$, the gas at a distance of $R \approx 5 \rt$ from the black hole has a density of about $\rho \approx 5 \times 10^{-10} \gcm3$ near the mid-plane while its radiation temperature is of order $T_{\rm r} \approx 5 \times 10^5 \kelvin$ according to Fig. \ref{fig:temperature_plane_relativistic}. These values yield a width of $H \approx 0.2 R = \rt$, which is consistent with that directly obtained from the simulation. Keeping the surface density fixed to $\Sigma = \rho H = 2\times 10^3 \gpercm2$ according to mass conservation, a disc in hydrostatic equilibrium with order unity aspect ratio would instead require a radiation temperature $T_{\rm r} \approx (3 G \mh \Sigma /a R^2)^{1/4} \approx 8 \times 10^5 \kelvin$, which is larger than anywhere in the simulation. The above calculations provide a simple justification for the small width resulting from fast radiative cooling, but they cannot account for the detailed disc properties that depend on additional processes. For instance, an additional source of disc confinement likely originates from the downward ram pressure gradient associated with the infalling matter continuously joining the disc from above and below the mid-plane. Although the disc possesses an overall thin profile, we additionally observe in Fig. \ref{fig:density_vertical_comparison} that the gas at its outer edges expands in the vertical direction as a result of heating by the infalling debris joining the circularized structure. This thickening appears to be more prominent on the side of the disc opposite to the intersection point for simulation A, which is likely because the injected matter preferentially reaches the mid-plane in this region.

\begin{figure}
\centering
\includegraphics[width=\columnwidth]{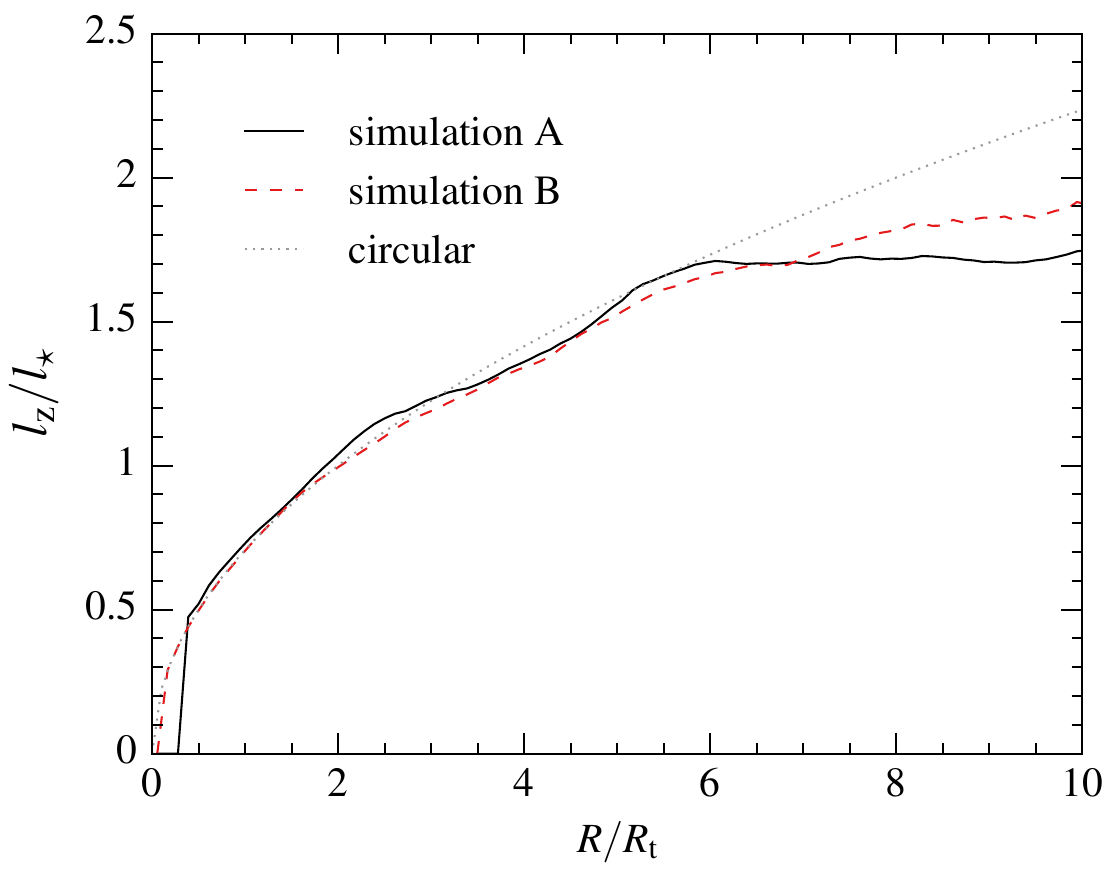}
\caption{Radial profile of the vertical component of the specific angular momentum vector evaluated at $t/\tmin = 0.4$ for simulation A (black solid line) and $t/\tmin = 0.5$ for simulation B (red dashed line) by considering gas contained within a vertical distance $|z|\leq \rt$ from the mid-plane where the formed disc is located. The grey dotted line represents the angular momentum $l_{\rm c} = (G \mh R)^{1/2}$ of a circular orbit in a Keplerian potential. Due to the units used, this curve is identical despite the different parameters considered in the two simulations.}
\label{fig:angmomvsr}
\end{figure}

The motion of the circularized debris can be further analysed from Fig. \ref{fig:angmomvsr} that shows the radial profile of the vertical component of its specific angular momentum for simulation A at $t/\tmin = 0.4$ (black solid line) and for simulation B at $t/\tmin = 0.5$ (red dashed line) by considering the gas contained within a vertical distance $|z| \leq 2 \rt$ from the mid-plane where the disc is located. These two profiles closely track the angular momentum $l_{\rm c} = \sqrt{G \mh R}$ corresponding to a circular orbit within a Keplerian potential (grey dotted line). The small deviation at large radii is due to the contribution from low angular momentum matter outside the disc, which is included because of the irregular boundary of the circularized gas. This near-Keplerian profile implies that the debris moves almost ballistically inside the disc, with the gravitational force from the black hole being entirely compensated by its centrifugal acceleration. This property results from the low pressure gradients of this gas due to efficient radiative cooling at secondary shocks. The dynamics of the disc is therefore different from that formed in our earlier adiabatic simulation \citep{bonnerot2020-realistic}, for which support against gravity was mostly provided by pressure.

\begin{figure}
\centering
\includegraphics[width=\columnwidth]{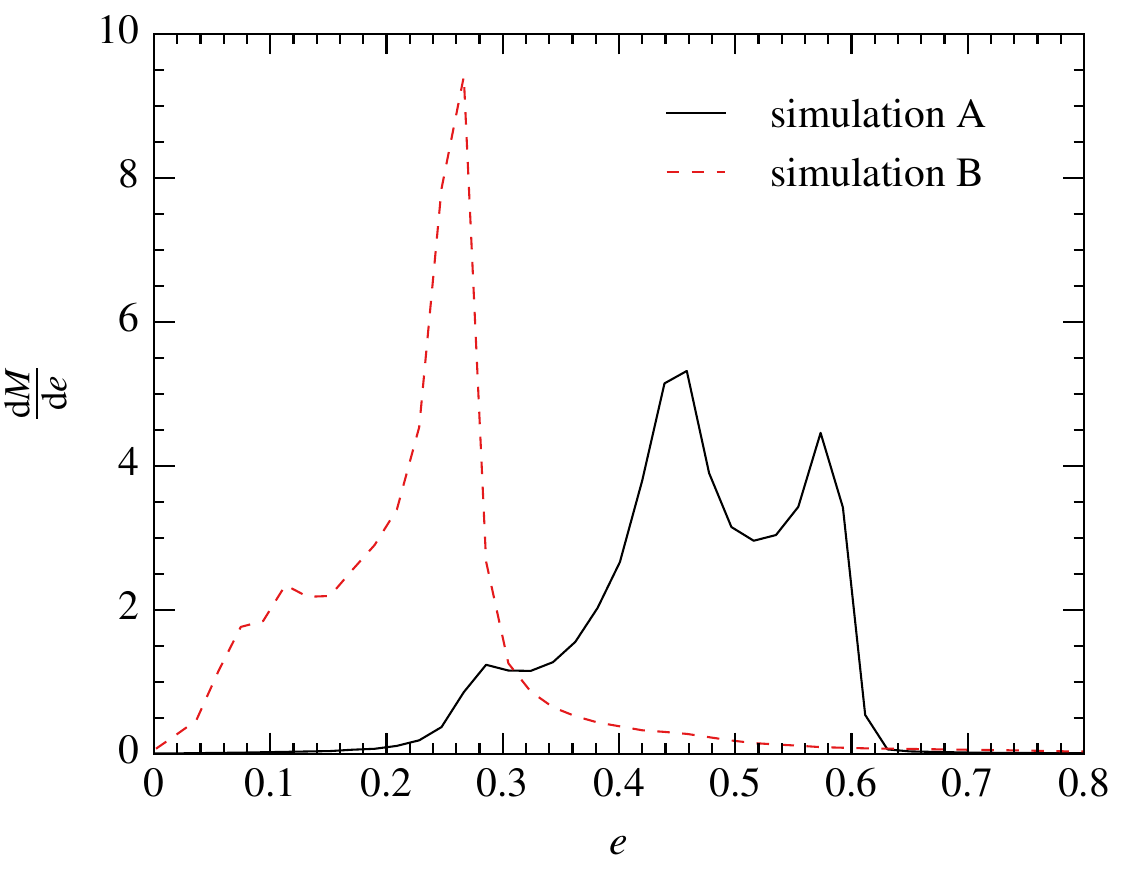}
\caption{Eccentricity distribution for simulation A at $t/\tmin = 0.4$ (black solid line) and simulation B at $t/\tmin = 0.5$ (red dashed line) considering the gas within a radius $R \leq 10 \rt$ and a vertical distance $|z|\leq 2 \rt$ from the mid-plane where the disc is located. The normalization is chosen such that the areas under the curves are equal to one.}
\label{fig:eccdist}
\end{figure}

Even after the disc has settled, the gas is not perfectly circular but retains instead significant eccentricities whose distribution is depicted in Fig. \ref{fig:eccdist} at $t/\tmin = 0.4$ for simulation A (black solid line) and at $t/\tmin = 0.5$ for simulation B (red dashed line). It is obtained from the gas located at $R\leq 10 \rt$ within a vertical distance $|z|\leq 2 \rt$ from the mid-plane. To calculate eccentricities, we make use of the Keplerian formula since the gas moves close to ballistically within the disc, as shown above. The normalization is chosen such that the area below the curves is equal to one. For simulation A, the debris has a broad range of eccentricities with an average value of $e \approx 0.5$. As visible at several times from Fig. \ref{fig:density_plane_relativistic}, this significant non-circularity is due to the lower extent of the disc towards the $\ey$ direction compared to the opposite side. We attribute this asymmetry to the continuous injection of linear momentum by the matter joining the disc from the nearby intersection point. Additionally, the distribution features several peaks of comparable heights that are caused by the presence of individual elliptical rings of higher density. The distribution for simulation B has a prominent peak at a lower average eccentricity of $e \approx 0.25$, which is likely caused by a reduced impact of the gas injected with lower speeds further away from the black hole. We emphasize that the exact gas eccentricities may be affected by relativistic apsidal precession, which is not considered in our Keplerian simulations. This effect would likely promote additional interactions between neighbouring trajectories, thus making the disc more circular overall.

\begin{figure}
\centering
\includegraphics[width=\columnwidth]{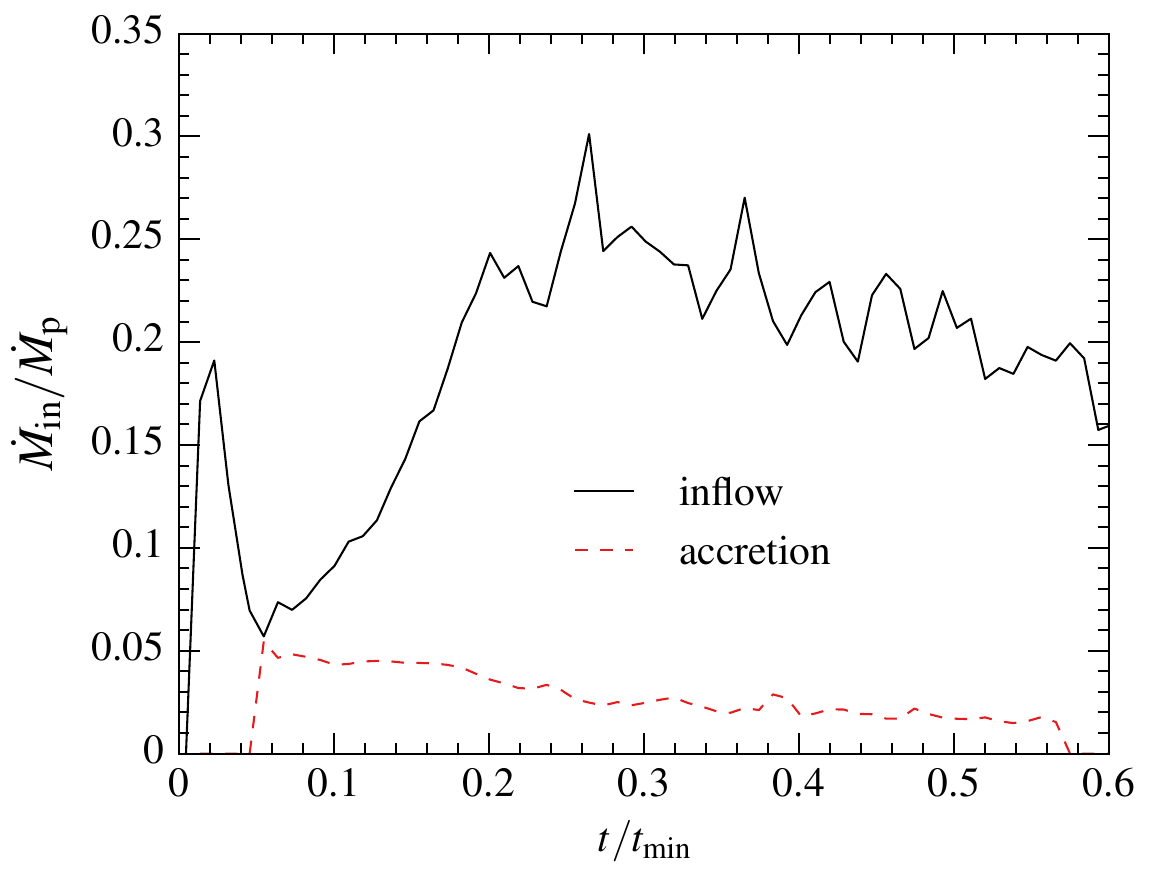}
\caption{Evolution of the net inflow rate for simulation A through a sphere of radius $R  = 20 \rt$ located outside the outer edge of the forming disc (black solid line) and the accretion rate onto the black hole (red dashed line) obtained from the particles crossing the accretion radius at $R_{\rm acc} = 6 \rg$.}
\label{fig:mdotinvst}
\end{figure}

A fraction of the injected debris proceeds to move towards the black hole where the accretion disc assembles. The evolution of the corresponding net inflow rate is shown in Fig. \ref{fig:mdotinvst} for simulation A by considering a sphere of radius $R = 20 \rt$ located close to the outer edge of the forming disc. It is computed from $\dot{M}_{\rm in} = (\Delta M_{\rm in} -\Delta M_{\rm out}) / \Delta t  $, where $\Delta M_{\rm out}$ and $\Delta M_{\rm in}$ are the masses of gas crossing the sphere during a time-step $\Delta t$ while moving inward and outward, respectively. The inflow rate features a prominent early peak at $t/\tmin \approx 0.03$ caused by the first arrival of gas in the region considered. Later on, it increases  as the disc builds up to reach $\dot{M}_{\rm in} \approx 0.25 \dot{M}_{\rm p} \approx 6 \dot{M}_{\rm Edd}$ by a time $t/\tmin \approx 0.2$. The rate remains constant at this value until the end of the simulation due to matter continuously joining the formed disc in a close to steady-state way. As illustrated with the two trajectories of gas particles shown in Fig. \ref{fig:velocity_plane_relativistic}, the gas arrives either directly from the intersection point (blue solid line) or from the $-\ey$ direction after striking the mid-plane at larger radii (orange dashed line). The red dashed line in Fig. \ref{fig:mdotinvst} indicates the rate of gas accretion onto the black hole that we determine from the particles entering the accretion radius at $R_{\rm acc} = 6 \rg$. It rapidly reaches its maximal value of $\dot{M}_{\rm acc} \approx 0.05 \dot{M}_{\rm p} = \dot{M}_{\rm Edd}$ before slowly decreasing until the end of the simulation. This accretion is due to almost ballistic gas trajectories that occasionally get deflected to enter the accretion radius. Inflow and accretion are qualitatively the same for simulation B, although with a delay due to the larger intersection radius. As expected from the low level of accretion compared to that of inflow, the disc mass increases at all times. We evaluate it as before from the gas located at $R\leq 10 \rt$ from the black hole and within a vertical distance $|z|\leq 2\rt$. At the end of simulations A and B at $t/\tmin = 0.6$, this mass reaches a similar value of $M_{\rm d} \approx 0.01 \msun$, which represents $20\%$ and $7\%$ of the total injected mass, respectively.

Since our simulation does not include magnetic fields, it does not capture the development of the magneto-rotational-instability \citetext{MRI, \citealt{balbus1991}} that likely results in enhanced accretion once it reaches saturation compared to the one we measure that is only caused by gas ballistically entering the accretion radius. Viscous accretion induced by these magneto-hydrodynamical effects would also cause additional heating compared to that associated with shocks only. Importantly, this may increase the gas internal energy close to the disc mid-plane while we have seen in Section \ref{sec:secondary} that dissipation by shocks occurs near the surface of the gas distribution at larger vertical heights. In this situation, we do not expect cooling to be as efficient due to the larger optical depths encountered by the radiation produced near the mid-plane. This is supported by Fig. \ref{fig:tauvsz}, which shows that the ratio of diffusion to dynamical times at $z = 0$ increases to $t_{\rm dif} / t_{\rm dyn} \approx \tau v/c \gg 1$. Once these additional dissipative processes start acting, it is therefore possible that the disc becomes significantly thicker than what we predict. We further discuss the influence of this additional physics in Section \ref{sec:accretion} but defer a detailed investigation to future works.

\begin{figure*}
\centering
\includegraphics[width=\textwidth]{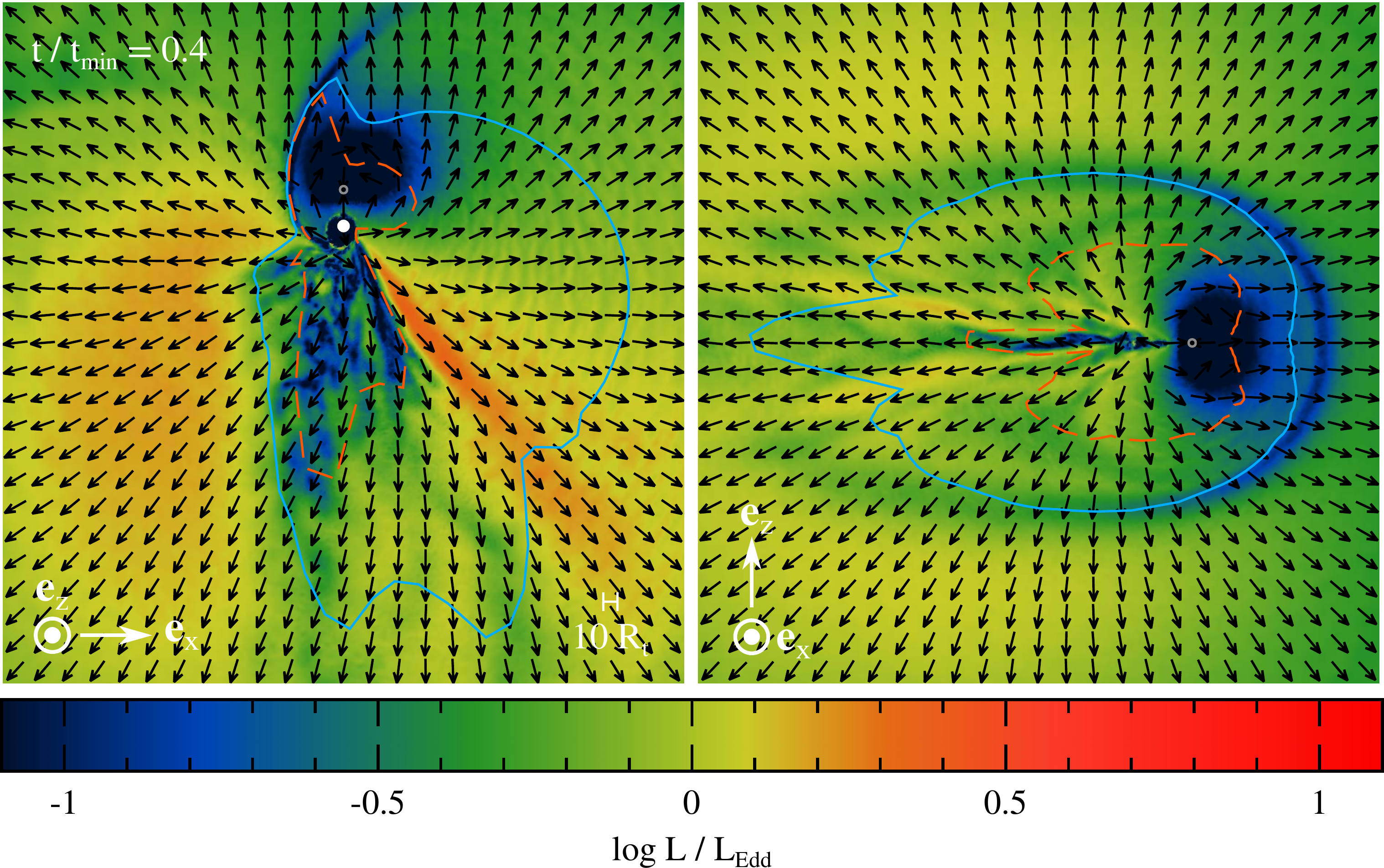}
\caption{Isotropic equivalent luminosity in slices parallel (left panel) and perpendicular (right panel) to the stellar orbital plane that contain the black hole for simulation A at $t/\tmin = 0.4$. It is computed from $L = 4 \pi R^2 (\vect{F}_{\rm r} \cdot \vect{e}_{\rm r})$ using the projection of the radiation flux directly obtained from the code along the radial direction, which is divided by the Eddington luminosity $L_{\rm Edd} = 4 \pi G \mh c / \kappa_{\rm s}\approx 4 \times 10^{44} \ergpers$. This ratio is also that of the accelerations due to radiation pressure and gravity along this direction. It increases from blue to red, as shown on the colour bar. The white segment indicates the scale used and the orientation is given by the white arrows. The black hole is depicted by the white dot and the grey circle represents the intersection point, from which matter is injected into the computational domain. The black arrows indicate the direction of the radiation flux while the blue solid lines denote the location of the scattering photosphere at the current time. The orange dashed curve denotes the estimated thermalization surface, where the properties of the emerging spectrum are set.}
\label{fig:luminosity_relativistic}
\end{figure*}

\begin{figure}
\centering
\includegraphics[width=\columnwidth]{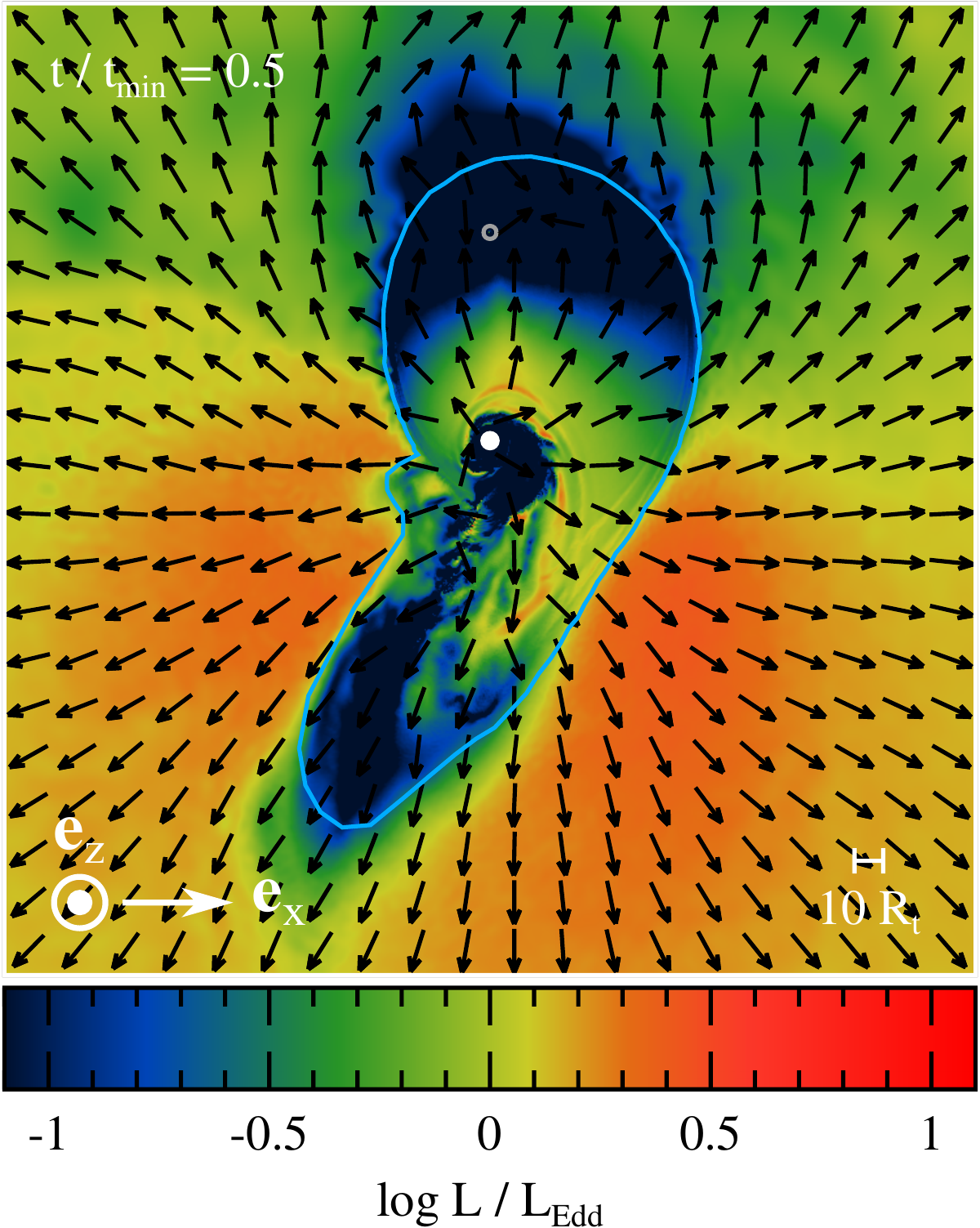}
\caption{Isotropic equivalent luminosity in a slice parallel to the stellar orbital plane that contain the black hole for simulation B at $t/\tmin = 0.5$. It is computed from $L = 4 \pi R^2 (\vect{F}_{\rm r} \cdot \vect{e}_{\rm r})$ using the projection of the radiation flux directly obtained from the code along the radial direction, which is divided by the Eddington luminosity $L_{\rm Edd} = 4 \pi G \mh c / \kappa_{\rm s} \approx 10^{44} \ergpers$. This ratio increases from blue to red, as shown on the colour bar. The white segment indicates the scale used and the orientation is given by the white arrows. The black hole is depicted by the white dot and the grey circle represents the intersection point, from which matter is injected into the computational domain. The black arrows indicate the direction of the radiation flux while the blue solid line denotes the location of the scattering photosphere at the current time.}
\label{fig:luminosity_plane_non_relativistic}
\end{figure}

\subsection{Irradiation of the debris and emerging luminosity}

\label{sec:emerging}

The radiation produced by secondary shocks proceeds to diffuse away from the forming disc to get transported to larger distances. As a result, the previously cold surrounding envelope gets irradiated, as can be seen in both simulations from Figs. \ref{fig:temperature_plane_relativistic} and \ref{fig:temperature_plane_non_relativistic} through a progressive increase of the radiation temperature that starts close to the black hole where it rises to $T_{\rm r} \approx 10^5 \kelvin$. Near the $-\ey$ direction, radiation propagates efficiently through gas of densities similar or lower than in the innermost regions. However, diffusion is slowed towards the intersection point where densities are enhanced due to the injection of debris from a small volume. Near this location, matter is entirely shielded from irradiation due to both larger optical depths that reduce the radiation flux and the fact that photons get additionally advected inward by the gas approaching the black hole. At a finite distance from the injection point, this inward advective transport compensates that due to diffusion and the net motion of the photons stalls, leaving the freshly-injected gas entirely free of radiation with $T_{\rm r} \ll 10^5 \kelvin$.

\begin{figure}
\centering
\includegraphics[width=\columnwidth]{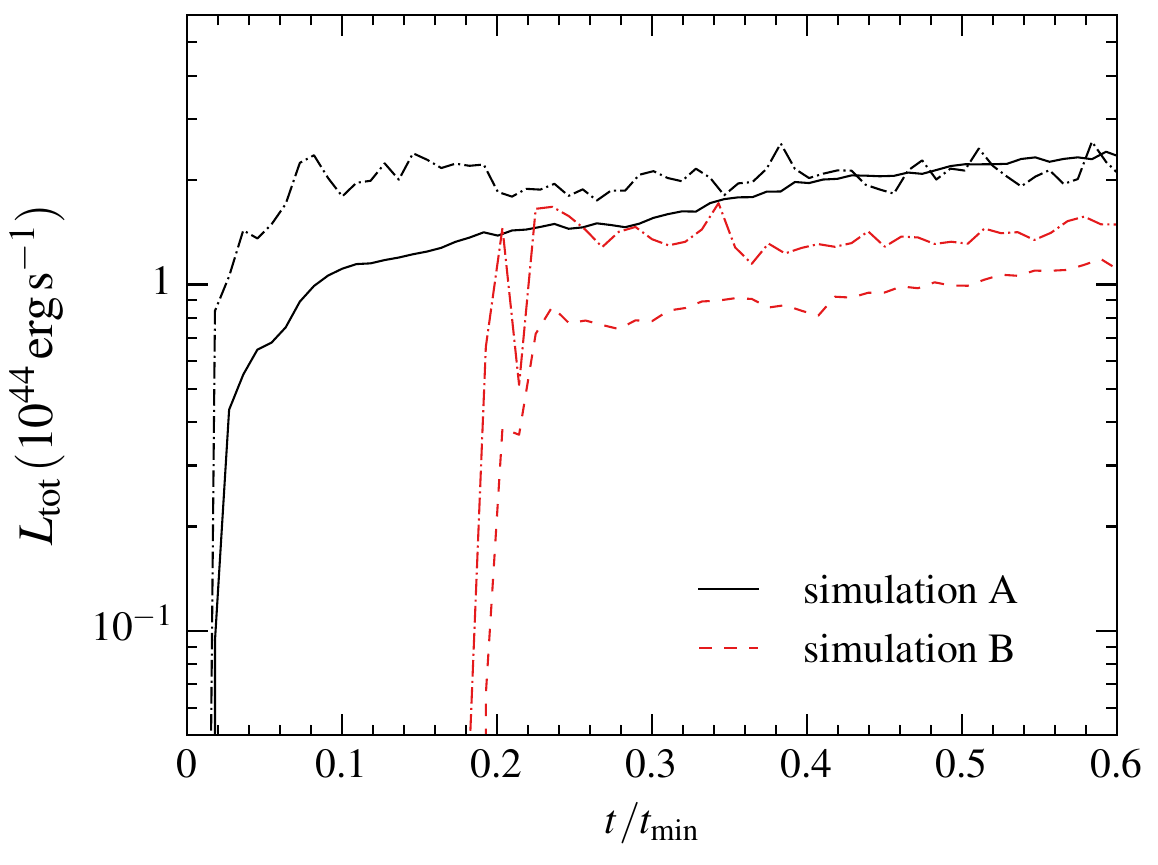}
\includegraphics[width=\columnwidth]{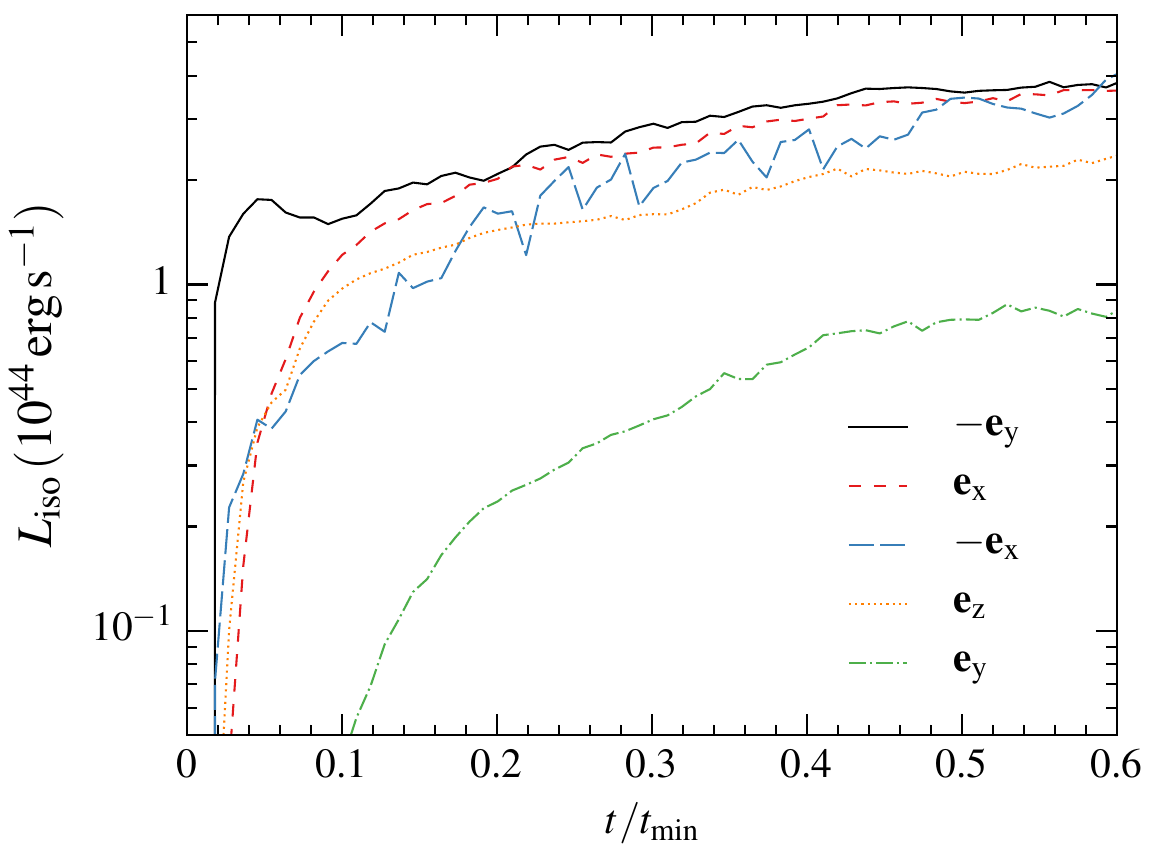}
\caption{Evolution of the total (upper panel) and isotropic equivalent (lower panel) luminosities. The total luminosity $L_{\rm tot}$ is shown for simulations A (black solid line) and B (red dashed line). It is obtained by integrating the radiation flux on the whole photosphere, which we evaluate at different times as explained in the text. These luminosities are compared to the corresponding heating rates indicated with dash-dotted lines of the same colours for both simulations. The isotropic equivalent luminosity $L_{\rm iso}$ is shown for simulation A along the $-\ey$ (solid black line), $\ex$ (red dashed line), $-\ex$ (blue long-dashed line), $\ez$ (orange dotted line) and $\ey$ (green dash-dotted line) directions. It is computed by integrating the radiation flux on the local photospheric surface by considering gas inside a small cone centred on the direction considered.}
\label{fig:lumvst}
\end{figure}

\begin{figure*}
\centering
\includegraphics[width=\textwidth]{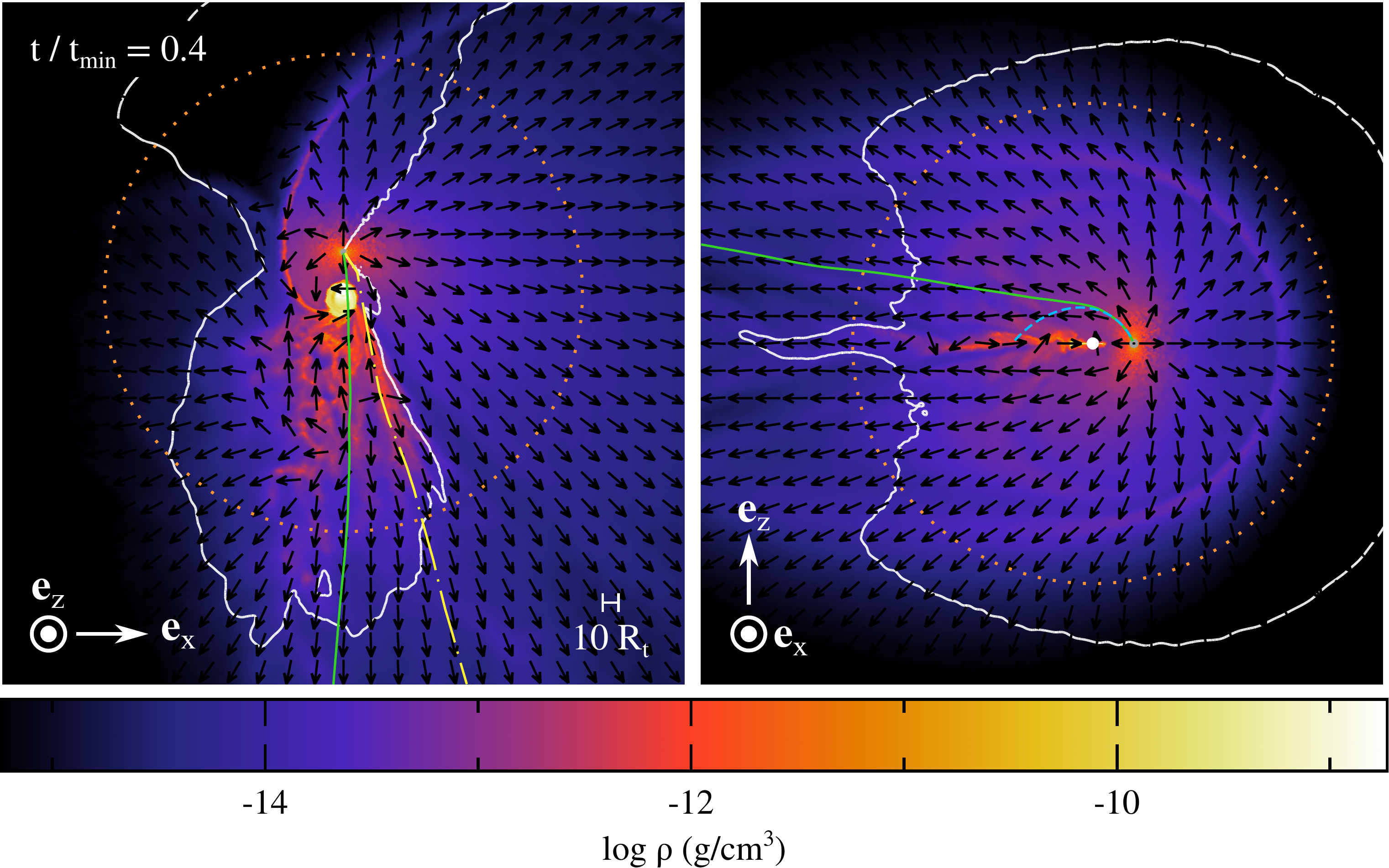}

\caption{Gas density shown on large scales for simulation A inside slices parallel (left panel) and orthogonal (right panel) to the stellar orbital plane that contain the black hole. The value of the density increases from black to white, as shown on the colour bar. The white segment indicates the scale used and the orientation is given by the white arrows. The black hole is depicted by the white dot and the grey circle represents the intersection point, from which matter is injected into the computational domain. The black arrows indicate the direction of the velocity field while the white contour corresponds to the surface beyond which the gas becomes unbound from the black hole with a positive orbital energy $\varepsilon = v^2/2 - G \mh /R >0$. The dotted orange circle located at $R=150 \rt$ indicates the spherical surface used to evaluate the outflow rate in Fig. \ref{fig:mdotoutvst}. The dash-dotted yellow line represents the trajectory of a gas particle injected on a parabolic orbit with $\epsilon =0$ at the lowermost edge of the cone of unbound matter launched from the intersection point. The solid green line shows the trajectory of an initially bound particle that gets ejected due to radiation pressure while it would follow the blue dashed line if it was moving on an entirely ballistic orbit.}
\label{fig:density_large_relativistic}
\end{figure*}

The intensity of diffusion on large scales can be evaluated from Fig. \ref{fig:luminosity_relativistic}, which shows the isotropic equivalent luminosity $L = 4 \pi R^2 (\vect{F}_{\rm r} \cdot \vect{e}_{\rm r})$ computed for simulation A at $t/\tmin = 0.4$ using the radiation flux directly obtained from the code that we normalize by the Eddington luminosity for the gas contained in a slice parallel (left panel) and orthogonal (right panel) to the stellar orbital plane. For simulation B, the same luminosity is presented in Fig. \ref{fig:luminosity_plane_non_relativistic} at a time $t/\tmin = 0.5$ by considering only the parallel slice. Because of the shielding effect described above, radiation does not efficiently diffuse near the intersection point, which creates a large shadow of reduced luminosity extending to the edge of the surrounding envelope. The luminosity is larger towards the $-\ey$ direction where its average value is only slightly lower than the Eddington value, i.e. $L \lesssim L_{\rm Edd}$. Radiation diffuses most efficiently through the matter outflowing to large distances directly from the injection point, particularly above and below the forming disc as well as inside the cone of unbound matter present in simulation A. In these regions, diffusion is almost radial, as indicated by the black arrows showing the direction of the radiation flux. As we show in Section \ref{sec:outflow}, this gas is subject to significant acceleration due to radiation pressure that can affect its dynamics. Due to its increased optical depths, we however find that the debris accumulated near the mid-plane remains largely opaque to the outgoing radiation.

The diffusing photons emerge from the stellar debris at the photosphere $R_{\rm ph}$ where the scattering optical depth integrated to infinity becomes unity, that is $\tau = \int_{R_{\rm ph}}^{\infty} \rho \kappa_{\rm s} \diff R = 1$. By numerically computing this integral along different radial paths from the gas density given by the simulations, we determine the location of the corresponding photosphere. At early times, we find that the photospheric surface coincides with the boundary of the injected envelope due to the compactness of this gas. In particular, this allows the photons produced by the first interactions taking place near the black hole to promptly emerge from the system along the $-\ey$ direction. This can for example be noticed in Fig. \ref{fig:temperature_plane_relativistic} at $t/\tmin = 0.03$ from an increased radiation temperature outside the injected matter.\footnote{Note that radiation can keep propagating outside the stellar debris due to the presence of low-mass bath particles that act as an optically thin surrounding medium, as described in Section \ref{sec:method}.} The photosphere expands at later times, especially towards the directions where the gas distribution is the most extended. Its location is indicated with blue solid curves in Figs. \ref{fig:luminosity_relativistic} and \ref{fig:luminosity_plane_non_relativistic} at the current times based on the slices considered. An increasing fraction of this growing photosphere simultaneously gets crossed by outgoing radiation, leading to the emergence of photons along a larger range of lines of sight. This process is slowed along the $\ey$ direction, where diffusion can only occur on curved paths around the injection point. A significant amount of radiation can nevertheless reach this part of the photosphere in simulation A, while it remains largely free of photons until the end of simulation B due to both larger densities and an increased intersection radius.

\begin{figure}
\centering
\includegraphics[width=\columnwidth]
{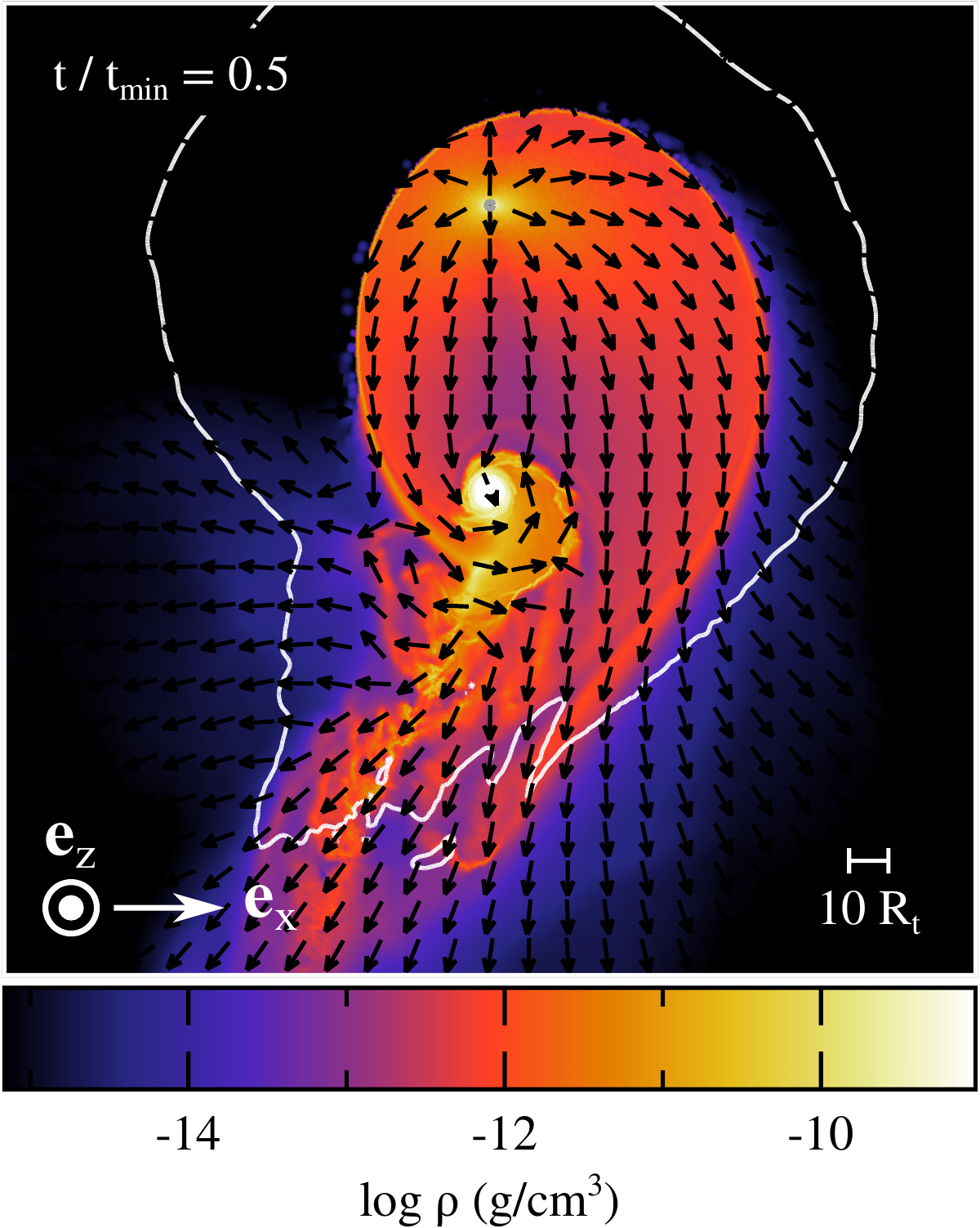}
\caption{Gas density shown on large scales for simulation B inside a slice parallel to the stellar orbital plane. The black hole is represented by the white dot while the white segment indicates the scale used.  The black arrows indicate the direction of the velocity field while the white contour corresponds to the surface beyond which the gas becomes unbound from the black hole with $\varepsilon = v^2/2 - G \mh /R >0$.}
\label{fig:density_large_plane_non_relativistic}
\end{figure}

\begin{figure}
\centering
\includegraphics[width=\columnwidth]{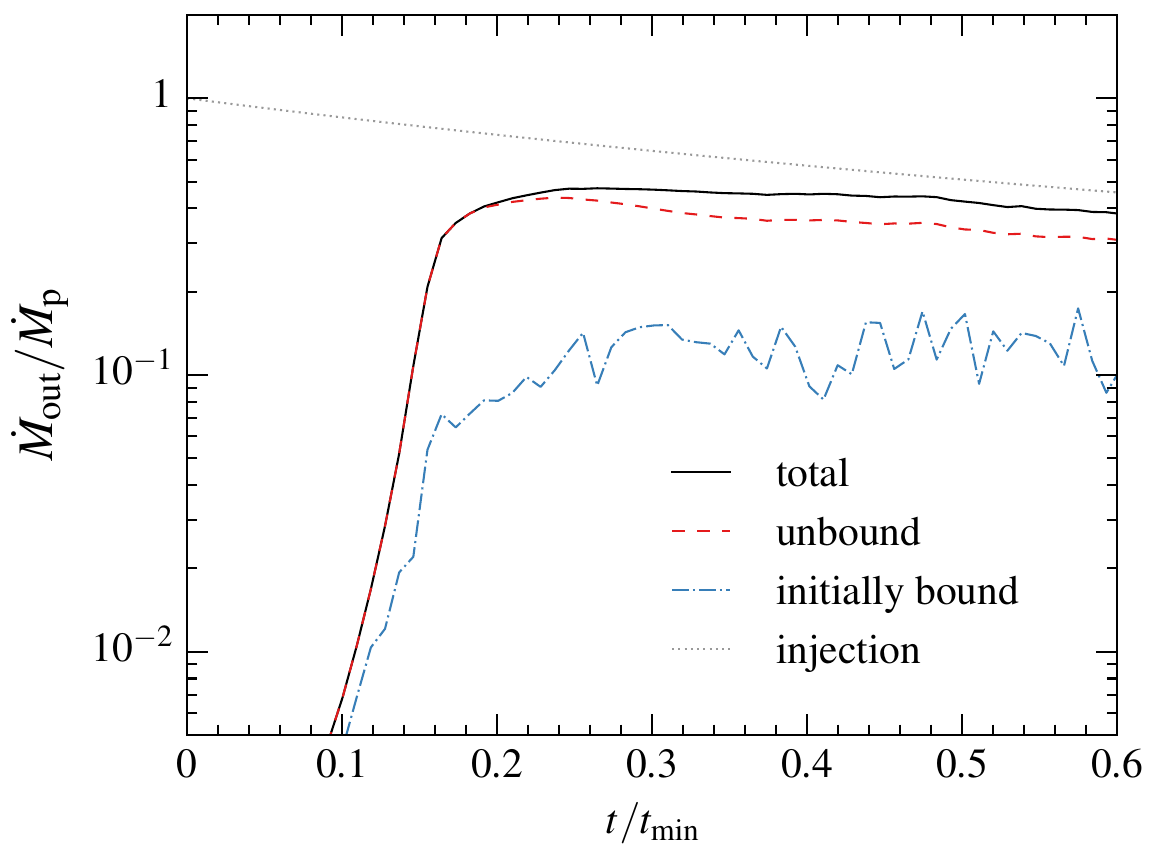}
\caption{Evolution of the total outflow rate (black solid line) for simulation A obtained from the gas crossing a spherical surface at $R=150 \rt$ while moving outward. The red dashed line indicates the unbound component, which is obtained by only considering the debris with positive orbital energy $\epsilon = v^2/2 - G \mh/R >0$. The contribution to this outflowing unbound gas from debris injected from the self-crossing shock on bound trajectories with $\epsilon < 0$ at injection is displayed with the blue dash-dotted line. The grey dotted line represents the rate of mass injection given by the fallback rate according to equation \eqref{eq:outflow_rate}.}
\label{fig:mdotoutvst}
\end{figure}

Radiation reaching the scattering photosphere emerges from the system, where it can keep propagating until potentially reaching an observer. The upper panel of Fig. \ref{fig:lumvst} shows the  evolution of the total luminosity obtained for simulations A (black solid line) and B (red dashed line), which are compared to the corresponding heating rates displayed with dash-dotted lines of the same colours. These luminosities are obtained by integrating the radiation flux on the whole three-dimensional photosphere determined above. In both simulations, the first shocks experienced by the infalling matter cause an early rise of the heating rate. Its short duration results from the small dynamical timescale near pericenter, which can be estimated as $t_{\rm p}= (\rp^3/G\mh)^{1/2} \approx 10^{-4} \tmin$. Due to the prompt emergence of photons produced by early interactions, this rise is closely tracked by the total luminosity. At later times, this luminosity keeps slowly increasing with a value of $L_{\rm tot} \approx 10^{44} \ergpers$ in the two simulations while remaining at all times slightly lower than the nearly constant heating rates. The fact that the emerging luminosity is similar to the heating rate implies that the radiation does not experience large adiabatic losses, which is consistent with the prompt diffusion taking place close to the dissipation sites.\footnote{As argued in the analytical work by \citet{metzger2016}, we expect adiabatic losses to be more significant for lower black hole masses due to a larger fallback rate, which can result in a lower peak of the lightcurve relative to the associated shock heating rate.} The evolution of isotropic equivalent luminosities is displayed for simulation A in the lower panel of Fig. \ref{fig:lumvst} for different lines of sight. Along the $-\ey$ (solid black line), $\ex$ (red dashed line), $-\ex$ (blue long-dashed line) and $\ez$ (orange dotted line) directions, these luminosities shows a prompt rise similar to that seen in the total luminosity. In addition, that along $-\ey$ features a peak associated with radiation from the early shocks described above, which is absent from the other curves. At later times, these luminosities keep slowly increasing from $L_{\rm iso} \approx 10^{44} \ergpers$ to values a few times larger. The luminosity along the $\ey$ direction (green dash-dotted line) displays a slower increase to eventually settle at a lower value of $L_{\rm iso} \lesssim 10^{44} \ergpers$. This qualitatively different evolution is due to the inefficient diffusion of radiation towards the intersection point. The corresponding diffusion timescale can be estimated using a density of $\rho \approx 10^{-12} \gcm3$ for the region surrounding the injection point as $t_{\rm diff} = \rint^2 \rho \kappa_{\rm s} / c \approx 0.06 \tmin$, which is consistent with the duration of the initial luminosity increase. For simulation B, we find that this reduction of the emerging luminosity is more pronounced as expected from the lower level of radiation in this region.

Even though our simulation does not directly provide the frequency distribution of the radiation evolved, we attempt to evaluate the emerging spectrum in an approximate manner from the radiative properties of the matter. In particular, we aim at roughly estimating the regions where the most energetic photons are absorbed, leading to reprocessed emission at lower energies, as proposed by several authors \citep{guillochon2014-10jh,metzger2016,roth2016}. In order to determine the ionization state of the gas, we compute the neutral fraction like \cite{metzger2016} from the balance between photoionization and recombination as $f_{\rm n} = 1/(1+ f_{\rm ion} c a T^4_{\rm r} \sigma_{\nu_{\rm ion}} / n_{\rm e} \alpha_{\rm B} h \nu_{\rm ion} )$, where $f_{\rm ion} = (\int_{\nu_{\rm ion}}^{+\infty} c e_{\nu} \sigma_{\nu} / h \nu \diff \nu) / (c a T^4_{\rm r} \sigma_{\nu_{\rm ion}} / h \nu_{\rm ion})$ specifies the fraction of ionizing radiation. Here, $e_{\nu}$ denotes the spectral energy density whose integrated value is $\int_{0}^{+\infty} e_{\nu} \diff \nu = e_{\rm r}$, the ionization cross-section is $\sigma_{\nu} = \sigma_{\nu_{\rm ion}} (\nu /\nu_{\rm ion})^{-3}$ with $\sigma_{\nu_{\rm ion}} = 6.304 \times 10^{-18} Z^{-2} \rm cm^2$ \citetext{equation (13.3) of \citealt{draine2011}}, and $\alpha_{\rm B} = 2.54 \times 10^{-13} Z^2 (T_{\rm g}/ Z^2 10^4 K)^{-0.8} \, \rm cm^3 s^{-1}$ \citetext{equation (14.6) of \citealt{draine2011}} represents the case B recombination coefficient. Following \citet{roth2016}, we only consider He II photo-ionization since it provides the main source of absorption for the gas properties of interest. Accordingly, the ionization potential is set to $h \nu_{\rm ion} = 54.4 \, \rm eV$  and the atomic number is $Z =2$. Given the neutral fraction, the bound-free opacity is determined from $\kappa_{\rm bf} = f_{\rm n} X \sigma_{\nu_{\rm ion}} / 2 Z m_{\rm p} $, using the solar value of $X=0.25$ for the Helium mass fraction. The typical temperature of the emerging spectrum is set at the thermalization radius $R_{\rm th}$ defined by $\int_{R_{\rm th}}^{\infty} \rho \kappa_{\rm eff} \diff R = 1$, where the effective opacity is given by $\kappa_{\rm eff} = (3\kappa_{\rm bf} (\kappa_{\rm bf}+ \kappa_{\rm s}))^{1/2}$ \citetext{equation (1.120) of \citealt{rybicki1979}}. This integrated optical depth takes into account the increased likelihood of absorption that results from the additional length covered by the photons due to scattering. By carrying out the above integral from outside in along radial paths, we determine the thermalization surface shown with the orange dashed line in Fig. \ref{fig:luminosity_relativistic} for simulation A, making use of a constant factor of $f_{\rm ion} = 0.03$.\footnote{Assuming local blackbody radiation, the ionization factor can numerically be evaluated from $f_{\rm ion} = (\int_{\nu_{\rm ion}}^{+\infty} 4 \pi B_{\nu} (T_{\rm r}) \sigma_{\nu} / h \nu \diff \nu) / (c a T^4_{\rm r} \sigma_{\nu_{\rm ion}} / h \nu_{\rm ion})$, where $B_{\nu}$ denotes the Planck function. This motivates our choice of $f_{\rm ion} = 0.03$ since it corresponds to a temperature of $T_{\rm r} = 8 \times 10^4 \kelvin$, which is the average value inside the gas located at the thermalization radius.} Most importantly, this surface extends towards the intersection point\footnote{Note that the thermalization surface also extends in the direction opposite to the intersection point, where it encloses the dense gas accumulated near the mid-plane due to its low radiation temperature. We highlight that this extended part is largely a consequence of the radial paths used to integrate the effective optical depth. In fact, radiation emerging around the $-\ey$ direction did not pass through the dense area but diffused around it after being produced at lower radii.} where it is crossed by gas with radiation temperature $T_{\rm r} \lesssim 3 \times 10^4 \kelvin$, as displayed in Fig. \ref{fig:temperature_plane_relativistic}.\footnote{While estimating the thermalization radius, we find that most of the gas is highly ionized such that the bound-free opacity is much lower than electron-scattering, i.e. $\kappa_{\rm bf}/\kappa_{\rm s} \ll 1$. This conclusion is consistent with equations (39) and (40) of the analytical work by \citet{metzger2016}, which considers gas properties and an energy injection similar to those of our simulations.} This suggests that radiation diffusing near the $\ey$ direction could primarily emerge as optical photons. Additionally, note that this emission could be bright with an isotropic equivalent luminosity reaching $L_{\rm iso} \approx 10^{44} \ergpers$ according to the green
dash-dotted line in the lower panel of Fig. \ref{fig:lumvst}. As we further discuss in Section \ref{sec:spectrum}, this effect may be at the origin of the early optical emission seen in TDEs, although this possibility would need to be further tested by means of detailed multi-frequency radiative transfer calculations.

Although photons are physically expected to move on close to straight lines directly outside the photosphere, our simulation finds instead that the direction of the radiation flux slightly varies. We attribute this unphysical behaviour to the FLD approach used in the simulation that becomes inappropriate in optically thin regions. As mentioned in Section \ref{sec:method}, the reason is that this technique imposes that radiation always flows towards the regions of lower radiation energy densities, even when the photons are free-streaming. A clear consequence is that the optically thin gas modelled with low-mass particles outside the distribution of injected debris gets irradiated even in the regions where little radiation is able to arrive at the photosphere, as can for example be seen from Fig. \ref{fig:luminosity_plane_non_relativistic} outside the envelope of debris in the direction of the self-crossing shock. This is because photons efficiently emerging from other parts of the injected gas move at close to the speed of light on curved paths that tend to rapidly fill the regions of low radiation content. However, this unphysical effect does not affect our estimate of luminosity since it is computed at the photospheric surface where the radiation is properly evolved. We also emphasize that our use of the FLD approximation is well justified inside the photosphere where most of the hydrodynamics described in this paper takes place.

\subsection{Outflow under radiation pressure}
\label{sec:outflow}

A fraction of the matter moves away from the black hole due to their trajectories at injection and additional deflections occurring during later interactions. The direction of the velocity field on large scales is indicated for simulation A with black arrows in Fig. \ref{fig:density_large_relativistic} that shows the gas density in slices parallel (left panel) and orthogonal (right panel) to the stellar orbital plane. This density distribution is also shown for simulation B in Fig. \ref{fig:density_large_plane_non_relativistic}, considering only the parallel slice. The white contour corresponds to the surface beyond which the gas has a positive orbital energy $\varepsilon = v^2/2 - G \mh/R >0$.\footnote{In Figs. \ref{fig:density_large_relativistic} and \ref{fig:density_large_plane_non_relativistic}, the white contour extends beyond the distribution of debris, which means that all the gas remains bound along this direction. This surface remains nevertheless artificially finite due to the presence of the low-mass bath particles filling the entire computational domain.} For simulation A, part of the outward motion is due to gas unbound due to its acceleration by the self-crossing shock. This matter is injected inside a cone aligned with the $\ex$ direction, whose boundary coincides on the left panel of Fig. \ref{fig:density_large_relativistic} with the white curve. The yellow dash-dotted line represents the trajectory of a fluid element injected at the lower edge of this cone near the stellar orbital plane on an initially parabolic orbit with $\varepsilon = 0$. The gravitational attraction from the black hole deflects this orbit, resulting in an increase of the solid angle occupied by the initially unbound debris that nevertheless remains confined to the right side of the intersection radius. This component is absent from simulation B since all the gas is injected on bound trajectories.

In both simulations, a large amount of initially bound matter moves outward near the $-\ey$ direction above and below the mid-plane. As described in Section \ref{sec:emerging}, this gas is continuously irradiated by the photons produced by secondary shocks close to the black hole, which is directly visible for simulation A from the right panel of Fig. \ref{fig:luminosity_relativistic}. The direction of the radiation flux  is close to perfectly aligned with the near-radial velocities of this gas, implying that it undergoes a positive work from the associated radiation pressure. In addition, the isotropic equivalent luminosity of $L \approx L_{\rm Edd}$ in this region implies that the corresponding acceleration largely compensates the gravitational pull from the black hole.\footnote{Equating the isotropic luminosity $L = 4 \pi R^2 (\vect{F}_{\rm r} \cdot \vect{e}_{\rm r})$ with the Eddington luminosity $L_{\rm Edd} = 4 \pi G \mh c / \kappa_{\rm s}$ implies $\kappa_{\rm s} (\vect{F}_{\rm r} \cdot \vect{e}_{\rm r}) / c \approx G \mh / R^2 $, which corresponds to a balance between the radiation pressure and gravitational accelerations along the radial direction (see equation \eqref{eq:momentum}).} This outflowing gas therefore experiences a continuous increase of orbital energy that can significantly modify its trajectories with a fraction eventually getting unbound. This unbinding is favoured by the low binding energy of the gas moving near the $-\ey$ direction, and results in a widening of the solid angle covered by unbound debris compared to that imparted by the self-crossing shock only. To analyse this effect, we follow the evolution of an individual gas particle in simulation A injected slightly above the mid-plane from the intersection point on a bound orbit. Assuming ballistic motion, this fluid element would follow the blue dashed line depicted in the right panel of Fig. \ref{fig:density_large_relativistic} that reaches a maximal vertical height before rapidly joining the mid-plane. In our simulation, it moves instead along the green solid line that reaches much larger distances due to the influence of radiation pressure until escaping the black hole gravity by crossing the white contour where $\epsilon = 0$.\footnote{Most of the unbound gas has a terminal velocity similar to the escape speed from the black hole of $v_{\rm esc} = (2 G \mh / R)^{1/2} = 0.03 c$ evaluated at $R = 150 \rt$. The outflow velocity can be much lower for the gas that stays bound since its radial motion cancels at apocenter.} As can be seen from Fig. \ref{fig:density_large_plane_non_relativistic}, part of this outflowing gas also gets unbound in simulation B but it represents a much lower contribution. In both simulations, the impact of radiation pressure on the trajectories is negligible for the gas located far away from the $-\ey$ direction either due to more bound trajectories that rapidly intersect near the black hole or because of a too low radiation content.

The evolution of the outflow rate is shown for simulation A with the black solid line in Fig. \ref{fig:mdotoutvst} for the gas crossing a sphere of radius $R = 150 \rt$. It is obtained from $\dot{M}_{\rm out} = \Delta M_{\rm out} / \Delta t$, where $\Delta M_{\rm out}$ is the mass of debris going outward through this spherical surface during a timestep $\Delta t$. This outflow rate rapidly rises when the debris reaches this distance either from the cone of unbound gas or along the $-\ey$ direction. Around $t/\tmin = 0.2$, it reaches its maximum value of $\dot{M}_{\rm out} \approx 0.5 \mdotp$ and then decreases until the end of the simulation. The outflow rate remains close to the injection rate shown with the grey dotted line, implying that the injected matter preferentially gets launched to large radii. A large fraction of this gas is unbound from the black hole, as can be seen from the red dashed line that displays the outflow rate only for the gas with a positive orbital energy $\epsilon >0$. This component is made of both gas belonging to the unbound cone and matter injected on bound trajectories that gained energy through radiation pressure, as described above. The dash-dotted blue line shows the contribution from this initially bound matter, which represents a significant fraction of the unbound outflow rate. We find a similar evolution for simulation B, although with a reduced outflow rate that is mostly made of bound gas.

\begin{figure}
\centering
\includegraphics[width=\columnwidth]{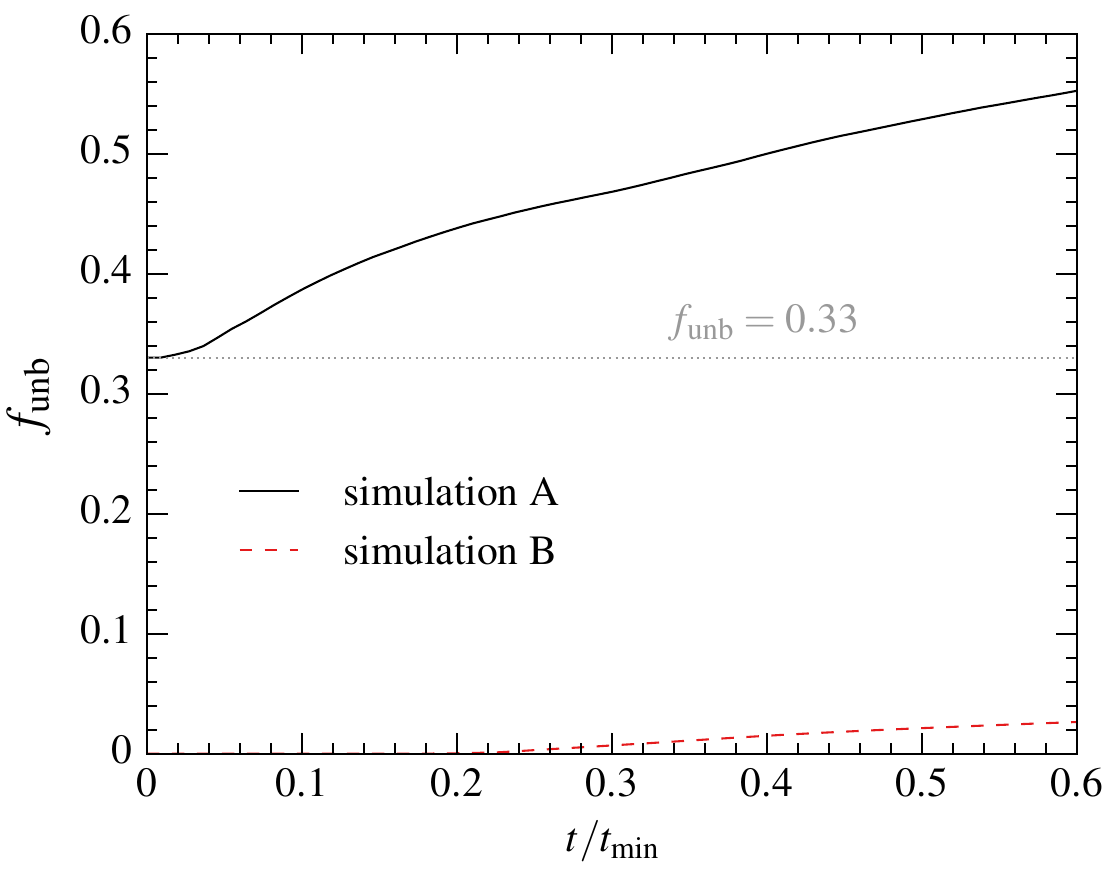}
\caption{Evolution of the unbound fraction for simulations A (black solid line) and B (red dashed line). It is calculated from ratio of the mass of unbound matter with a positive orbital energy $\epsilon = v^2/2 - G \mh/R >0$ to that of all the evolving gas. The horizontal grey dotted line denotes the unbound fraction $f_{\rm unb} = 0.33$ of the gas launched from the self-crossing shock in simulation A. For simulation B, all the gas is injected on bound orbits.}
\label{fig:funbvst}
\end{figure}

Fig. \ref{fig:funbvst} shows the evolution of the fraction of unbound matter obtained by considering all the injected gas for simulations A (black solid line) and B (red dashed line). Initially, the unbound fraction for simulation A is that $f_{\rm unb} = 0.33$ of the injected matter, which is depicted by the horizontal grey dotted line. Due to additional unbinding induced by radiation pressure, it then increases at all times to reach $f_{\rm unb} = 0.55$ at the end of the simulation. Despite a similar energy injected by shocks, this increase of the unbound fraction was not present in our earlier study \citep{bonnerot2020-realistic}. This is because this internal energy was confined near the black hole where gas is the most bound due to the assumption of adiabaticity. In the study presented here, we find that radiative diffusion allows energy to get transported to larger distances where less bound gas is located. This matter can more easily get unbound due to radiation pressure, which results in a larger fraction of gas able to escape the system. As expected, the unbound fraction is much lower for simulation B because all the gas is injected on bound orbits. It also experiences a smaller increase to reach a value of only $f_{\rm unb} = 0.02$ at the end of the simulation. This is due to the fact that the gas moves on average along more bound orbits after passing through the self-crossing shock. As a result, the increase of orbital energy results in less unbound matter than in simulation B, even though the energy dissipation through secondary shocks is similar, according to the dash-dotted lines in the upper panel of Fig. \ref{fig:lumvst}.

\section{Discussion and conclusion}

\label{sec:discussion}

\subsection{Treatment of the self-crossing shock}

\label{sec:self}

We treat the self-crossing shock by injecting gas inside the computational domain according to the properties of the resulting outflow determined from a simulation of this initial interaction by \cite{lu2020}. We emphasize that the strength of this collision was probably maximized in this local study by assuming that the two stream components are identical such that all the incoming gas gets efficiently shocked. In this situation, we found that the interaction most often leads to the formation of a quasi-spherical outflow, which is used as initial conditions of the disc formation simulation presented here. However, it is possible that the self-crossing shock is less efficient than we assume due to additional physical effects not considered. Nodal precession occurring around spinning black holes can cause the two incoming streams to move on different orbital planes, resulting in an offset that prevents a fraction of the gas from colliding. This effect was included in simulations of disc assembly \citep{hayasaki2016-spin,liptai2019-spin} but the bound stellar orbits they considered for simplicity makes the results difficult to extrapolate to the more likely case of a parabolic trajectory. Additionally, if the infalling stream is heated non-adiabatically during its pericenter passage due to the so-called nozzle shock, the component moving away from the black hole may become thicker than that still approaching such that some debris could miss the initial intersection. Analytical arguments \citep{guillochon2014-10jh} show that its influence grows with decreasing black hole mass, suggesting that it strongly affects the overall gas dynamics only for masses $\mh \lesssim 10^5 \msun$ lower than we consider. It is nevertheless possible that more modest changes of trajectories induced by the nozzle shock for a wider region of parameter space have a significant impact on the subsequent stream intersection. Settling this issue would require a systematic numerical study of the stream passage at pericenter performed at sufficiently high resolution, which we defer to the future. 

Another implicit assumption of our numerical treatment is that the debris stream continuously falling back to pericenter  evolves independently from the gas that has already passed through the self-crossing shock. As mentioned in our previous study \citep{bonnerot2020-realistic}, the stream is generally much denser than the shocked gas interacting near the black hole, which suggests that its trajectory remains largely unaffected by its passage through the forming disc. However, the stream may additionally be affected by hydrodynamical instabilities similar to those investigated in the presence of an external gaseous medium \citep{bonnerot2016-kh,kathirgamaraju2017}. A recent simulation by \citet{andalman2020} finds that such stream-disc interactions become rapidly dominant over the self-crossing shock. However, the deep disruption with $\beta = 7$ they consider may tend to favour this evolution due to an enhanced expansion of the stream at pericenter that makes it more susceptible to interactions with the surrounding matter. Even if the stream remains unaffected, its passage through the disc may also affect the evolution of the circularizing gas. One possible consequence is a reduction of the disc angular momentum, which could enhance accretion onto the black hole. Improving our understanding of the interplay between these different gaseous components would require additional numerical works beyond the scope of this paper.

As mentioned in Section \ref{sec:initial}, the self-crossing shock is assumed to be continuous in our simulations while it is in fact intermittent, since the outgoing stream component needs to get replenished after it has entirely passed through the intersection point. This implies that the outflow launched by the self-crossing shock is expected to be switched on and off periodically according to the replenishment timescale, given by the duration necessary for an element of stream to go from the intersection point to the black hole and back, which can be estimated as $t_{\rm rep} \approx 2(\rint^3/G \mh)^{1/2}$. If the self-crossing shock occurs relatively close to pericenter like in simulation A, this timescale is $t_{\rm rep}\ll \tmin$, which implies that the outflowing gas is made of thin shells injected very close to each other. We expect these shells to promptly merge together as they expand due internal pressure gradients, such that the dynamics is unlikely to differ from that assuming a continuous mass injection. However, if the intersection radius approaches the apocenter of the stream like in simulation B where $\rint \approx 1.4 \amin$, the replenishment timescale becomes $t_{\rm rep} \approx \tmin$, such that the gas launched from the self-crossing shock is made of thick and widely separated shells. In this case, the resulting emission could be modulated due to intermittent heating by secondary shocks near the black hole.

\subsection{Consequences on the early emission}

\label{sec:spectrum}

Our simulations show that radiative diffusion towards the intersection point is hampered due enhanced densities and the opposite advection of radiation, which results in a shallower increase of the lightcurve along this direction (green dot-dashed line in the lower panel of Fig. \ref{fig:lumvst}). Although this effect is directly caused by our treatment of the self-crossing shock discussed in Section \ref{sec:self}, we expect it to remain present or even strengthened if this interaction is taken into account more accurately. While we assume that all the incoming stream collides in a small region of space, a less idealized intersection would likely result in dense matter spreading a larger volume around the intersection point, thus further preventing efficient diffusion. On the opposite side, we find that photons are instead able to promptly escape such that the associated signal may more directly track the innermost interactions experienced by the circularizing gas. However, the sharp rise of the bolometric lightcurve to its maximal value is largely due to our assumption of a fallback rate peaking at $t=0$, made for numerical convenience. More realistically, the stellar structure imposes that this rate of arrival rises first to a peak value \citep{lodato2009,guillochon2013}. This would likely result in a slower increase of the lightcurve that tracks this early fallback rate.

The properties of the emerging spectrum are set at the thermalization surface, which we estimated as displayed with the orange dashed line in Fig. \ref{fig:luminosity_relativistic}. Near the intersection point, radiation diffuses through this surface with a temperature $T_{\rm r} \lesssim 3 \times 10^4 \kelvin$ much lower than near the black hole, which suggests that optical photons may emerge from this region. Such a reprocessing mechanism for the origin of optical emission in TDEs has been studied in several works \citep{guillochon2014-10jh,metzger2016,roth2016}, but often relying on spherical symmetry. Under this assumption, they find that the reprocessing envelope needs to be of order a solar mass \citep{roth2016} or more \citep{matsumoto2020} to reproduce the optical emission observed. The possibility of a higher reprocessing efficiency near the intersection point could alleviate this constraint by decreasing the required mass of debris. This would lend further support to this mechanism, by making it more generally applicable to encounters involving low-mass stars and partial disruptions where the mass reaching the black hole is reduced. Confirming this impact on the emerging spectrum would require detailed multi-frequency radiative transfer calculations through the Monte Carlo method, which we defer to future works.

\subsection{Subsequent phase of accretion}

\label{sec:accretion}

As the disc assembles, gas may proceed to accrete onto the black hole due to viscous effects driven by the gas magnetic fields, which are usually thought to start acting after the MRI has developed. The work by  \cite{sadowski2016} finds that magnetic fields only have a small impact on the gas evolution, although considering a deep disruption with $\beta = 10$ with the assumption of gas adiabaticity. Because magnetic fields are not included, this effect cannot be captured in our simulations and the low level of accretion that takes place is entirely ballistic. We nevertheless emphasize that the thin vertical profile of the disc may increase its viscous timescale, thus delaying the onset of gas accretion. However, an alternative possibility it that early dissipation associated with viscous effects leads to significant internal energy injection near the mid-plane. As mentioned in Section \ref{sec:circularization}, diffusion may be inefficient from this region of high density such that the resulting pressure gradients could promptly puff up the disc. Additionally, such prompt accretion may result in the launching of powerful winds and potentially a relativistic jet that can strongly modify the properties of the surrounding gas and radiation. If the black hole has a high enough spin, Lense-Thirring torques may also add some level of precession to the nascent accretion disc \citep[e.g.][]{stone2012}.

Gas accretion is expected to dominate at later times, after most of the bound debris has fallen back near the black hole. This stage of evolution has been studied by several numerical works \citep{dai2018,curd2018}, which find that accretion produces fast outflows possibly accompanied by a relativistic jet. For numerical convenience, these magneto-hydrodynamics simulations are initialized with a compact and isolated torus, but it is so far unclear whether these properties are appropriate for a TDE. Understanding the dynamics of disc formation is critical to determine the gas distribution present around the black hole once accretion starts. For example, our simulations suggest that the formed disc only contains a fraction of the returning debris while the rest forms a massive envelope at larger distances. It is possible that this configuration leads to an interplay between matter launched from the disc and the envelope, which could have observational consequences. Accurately studying the onset and subsequent impact of accretion processes requires radiation magneto-hydrodynamical simulations, which will be performed in the future.

\section{Summary}

\label{sec:summary}

We have carried out the first radiation-hydrodynamics simulations of accretion disc formation in a TDE. Our treatment of radiative processes relies on the FLD approximation, which is appropriate given the large gas optical depths involved. Like in our previous study \citep{bonnerot2020-realistic} assuming gas adiabaticity, the self-crossing shock is treated through an injection of outflowing matter into the computational domain, which allows us to consider realistic parameters of the problem. The same initial conditions are used in simulation A as in our earlier investigation that corresponds to a significantly relativistic encounter. In simulation B, the parameters are modified such that relativistic effects are reduced, leading to a weaker self-crossing shock that takes place further away from the black hole. The results of this investigation can be summarized as follows.

\begin{enumerate}

\item The injected debris moves on a wide range of trajectories that intersect near the black hole, causing the formation of secondary shocks. Free-free emission from the hot gas promptly leads to an increase of the radiation energy with a corresponding temperature reaching $T_{\rm r} \approx 10^5 \kelvin$ in the innermost regions.

\item Optical depths are low enough for the deposited internal energy to promptly diffuse away from the interacting gas in the form of radiation. Owing to this efficient cooling, the infalling matter accumulates into a thin sheet near the stellar orbital plane that extends in the direction opposite to the injection point.

\item Secondary shocks lead to the formation of an accretion disc that starts assembling shortly after the arrival of the first debris. Infalling gas continuously joins the nascent disc such that its mass keeps increasing to reach $M_{\rm d} \approx 0.01 \msun$ at the end of the simulations, which represents only a small fraction of all the injected matter.

\item The accretion disc extends to a distance $R \approx 10 \rt$ from the black hole. It remains thin with an approximately uniform width of $H \approx 2 \rt$ and features a near-Keplerian angular momentum profile. This results from an efficient radiative cooling at circularizing shocks that prevents the gas internal energy from building up, thus reducing the associated pressure gradients.

\item In both simulations, the formed disc rotates in the direction opposite to that of the star around the black hole due to the angular momentum sign of the most bound injected debris. Even after the disc properties have settled, the gas within it retains moderately large eccentricities inside the range $0<e\lesssim 0.6$ with significantly larger values for simulation A.

\item Radiative diffusion leads to the irradiation of the previously cold envelope of gas surrounding the forming accretion disc. Diffusion is most efficient within the gas outflowing on the side opposite to the intersection point where the isotropic equivalent luminosity reaches $L \approx L_{\rm Edd}$. In the direction of the self-crossing shock, diffusion is however slowed due to higher densities and an opposite advective flux, which results in a much lower radiation content.

\item Radiation escapes from the gas when it reaches the scattering photosphere, which continuously grows as the injected matter reaches larger distances. The total luminosity leaving the system features a sharp rise shortly after the first secondary shocks occur. At later times, it slowly increases with a value of $L_{\rm tot} \approx 10^{44} \ergpers$ in both simulations. This luminosity is only slightly lower than the corresponding heating rates, which implies small adiabatic losses.

\item Along most directions, the emerging isotropic equivalent luminosity promptly reaches $L_{\rm iso} \approx 10^{44} \ergpers$ to then increase by a factor of a few until the end of  the simulations. Due to inefficient diffusion, this increase is much slower towards the intersection point. The corresponding luminosity nevertheless reaches a significant value of $L_{\rm iso} \lesssim 10^{44} \ergpers$ for simulation A while it remains much lower for simulation B because of a larger intersection radius.

\item The thermalization surface estimated from the gas properties taking into account bound-free opacity is crossed by diffusing radiation with temperature $T_{\rm r} \lesssim 3 \times 10^4 \kelvin$ near the intersection point. This suggests that optical photons may emerge from this region as found observationally, although this mechanism needs to be further tested with more detailed calculations.

\item A large fraction of the injected gas gets to large distances either directly from the intersection radius or after experiencing interactions near the black hole. Most of it is irradiated at a near-Eddington luminosity, which causes an increase of orbital energy that can eventually unbind some of this gas. For simulation A, the fraction of unbound gas increases from that $f_{\rm unb} = 0.33$ caused by the self-crossing shock only to a significantly larger final value of $f_{\rm unb} = 0.55$. This fraction remains however small for simulation B due to the more bound trajectories of the injected debris.

\end{enumerate}

This study paves the way to a detailed understanding of disc formation and the radiation produced during this early phase of evolution that represents the first observable light from a TDE. We intend to pursue this effort, particularly by evaluating the influence of magnetic fields on the hydrodynamics and the most salient properties of the emerging electromagnetic spectrum. Significant improvements are expected in the theoretical characterization of this early emission, which is now accessible by many observations.

\section*{Acknowledgments}

We thank Tony Piro, Eliot Quataert, Sterl Phinney, Daniel Kasen and Nathan Roth for useful discussions. We acknowledge the use of \textsc{splash} \citep{price2007} for producing most of the figures in this paper. This research benefited from interactions at the ZTF Theory Network Meeting, partly funded by the National Science Foundation under Grant No. NSF PHY-1748958. The research of CB was funded by the Gordon and Betty Moore Foundation through Grant GBMF5076. WL was supported by the David and Ellen Lee Fellowship at Caltech.

\section*{Data availability}

The data underlying this article will be shared on reasonable request to the corresponding author. A public version of
the GIZMO code is available at \url{http://www.tapir.caltech.edu/~phopkins/Site/GIZMO.html}.




\bibliographystyle{mnras} 
\bibliography{biblio}

\begin{thebibliography}{}
\makeatletter
\relax
\def\mn@urlcharsother{\let\do\@makeother \do\$\do\&\do\#\do\^\do\_\do\%\do\~}
\def\mn@doi{\begingroup\mn@urlcharsother \@ifnextchar [ {\mn@doi@}
  {\mn@doi@[]}}
\def\mn@doi@[#1]#2{\def\@tempa{#1}\ifx\@tempa\@empty \href
  {http://dx.doi.org/#2} {doi:#2}\else \href {http://dx.doi.org/#2} {#1}\fi
  \endgroup}
\def\mn@eprint#1#2{\mn@eprint@#1:#2::\@nil}
\def\mn@eprint@arXiv#1{\href {http://arxiv.org/abs/#1} {{\tt arXiv:#1}}}
\def\mn@eprint@dblp#1{\href {http://dblp.uni-trier.de/rec/bibtex/#1.xml}
  {dblp:#1}}
\def\mn@eprint@#1:#2:#3:#4\@nil{\def\@tempa {#1}\def\@tempb {#2}\def\@tempc
  {#3}\ifx \@tempc \@empty \let \@tempc \@tempb \let \@tempb \@tempa \fi \ifx
  \@tempb \@empty \def\@tempb {arXiv}\fi \@ifundefined
  {mn@eprint@\@tempb}{\@tempb:\@tempc}{\expandafter \expandafter \csname
  mn@eprint@\@tempb\endcsname \expandafter{\@tempc}}}

\bibitem[\protect\citeauthoryear{Andalman, Liska, Tchekhovskoy, Coughlin  \&
  Stone}{Andalman et~al.}{2020}]{andalman2020}
Andalman Z.~L.,  Liska M. T.~P.,  Tchekhovskoy A.,  Coughlin E.~R.,   Stone N.,
   2020, preprint, 20, 1 (\mn@eprint {arXiv} {2008.04922})

\bibitem[\protect\citeauthoryear{Arcavi et~al.,}{Arcavi
  et~al.}{2014}]{arcavi2014}
Arcavi I.,  et~al., 2014, \mn@doi [\apj] {10.1088/0004-637X/793/1/38}, 793, 38

\bibitem[\protect\citeauthoryear{Balbus \& Hawley}{Balbus \&
  Hawley}{1991}]{balbus1991}
Balbus S.~A.,  Hawley J.~F.,  1991, \mn@doi [\apj] {10.1086/170270}, 376, 214

\bibitem[\protect\citeauthoryear{Blagorodnova et~al.,}{Blagorodnova
  et~al.}{2017}]{blagorodnova2017}
Blagorodnova N.,  et~al., 2017, \apj, 1, 1

\bibitem[\protect\citeauthoryear{Blagorodnova et~al.,}{Blagorodnova
  et~al.}{2019}]{blagorodnova2019}
Blagorodnova N.,  et~al., 2019, \mn@doi [\apj] {10.3847/1538-4357/ab04b0}, 873,
  92

\bibitem[\protect\citeauthoryear{Bonnerot \& Lu}{Bonnerot \&
  Lu}{2020}]{bonnerot2020-realistic}
Bonnerot C.,  Lu W.,  2020, \mn@doi [\mnras] {10.1093/mnras/staa1246}, 495,
  1374

\bibitem[\protect\citeauthoryear{Bonnerot \& Stone}{Bonnerot \&
  Stone}{2020}]{bonnerot2020-review}
Bonnerot C.,  Stone N.,  2020, preprint, 91125 (\mn@eprint {arXiv}
  {2008.11731})

\bibitem[\protect\citeauthoryear{Bonnerot, Rossi, Lodato  \& Price}{Bonnerot
  et~al.}{2016a}]{bonnerot2016-circ}
Bonnerot C.,  Rossi E.~M.,  Lodato G.,   Price D.~J.,  2016a, \mn@doi [\mnras]
  {10.1093/mnras/stv2411}, 455, 2253

\bibitem[\protect\citeauthoryear{Bonnerot, Rossi  \& Lodato}{Bonnerot
  et~al.}{2016b}]{bonnerot2016-kh}
Bonnerot C.,  Rossi E.~M.,   Lodato G.,  2016b, \mn@doi [\mnras]
  {10.1093/mnras/stw486}, 458, 3324

\bibitem[\protect\citeauthoryear{Bonnerot, Rossi  \& Lodato}{Bonnerot
  et~al.}{2017a}]{bonnerot2017-stream}
Bonnerot C.,  Rossi E.~M.,   Lodato G.,  2017a, \mn@doi [\mnras]
  {10.1093/mnras/stw2547}, 15, 1

\bibitem[\protect\citeauthoryear{Bonnerot, Price, Lodato  \& Rossi}{Bonnerot
  et~al.}{2017b}]{bonnerot2017-magnetic}
Bonnerot C.,  Price D.~J.,  Lodato G.,   Rossi E.~M.,  2017b, \mn@doi [\mnras]
  {10.1093/mnras/stx1210}, 469, 4879

\bibitem[\protect\citeauthoryear{Brassart \& Luminet}{Brassart \&
  Luminet}{2010}]{brassart2010}
Brassart M.,  Luminet J.-P.,  2010, \mn@doi [\aap]
  {10.1051/0004-6361/200913442}, 511, A80

\bibitem[\protect\citeauthoryear{Bricman \& Gomboc}{Bricman \&
  Gomboc}{2019}]{Bricman2020}
Bricman K.,  Gomboc A.,  2019, preprint (\mn@eprint {arXiv} {1906.08235})

\bibitem[\protect\citeauthoryear{Carter \& Luminet}{Carter \&
  Luminet}{1982}]{carter1982}
Carter B.,  Luminet J.~P.,  1982, \mn@doi [Nature] {10.1038/296211a0}, 296, 211

\bibitem[\protect\citeauthoryear{Chatzopoulos \& Weide}{Chatzopoulos \&
  Weide}{2019}]{chatzopoulos2019}
Chatzopoulos E.,  Weide K.,  2019, \mn@doi [\apj] {10.3847/1538-4357/ab18f9},
  876, 148

\bibitem[\protect\citeauthoryear{Chornock et~al.,}{Chornock
  et~al.}{2014}]{chornock2014}
Chornock R.,  et~al., 2014, \mn@doi [\apj] {10.1088/0004-637X/780/1/44}, 780,
  44

\bibitem[\protect\citeauthoryear{Commer{\c{c}}on, Teyssier, Audit, Hennebelle
  \& Chabrier}{Commer{\c{c}}on et~al.}{2011}]{commercon2011}
Commer{\c{c}}on B.,  Teyssier R.,  Audit E.,  Hennebelle P.,   Chabrier G.,
  2011, \mn@doi [\aap] {10.1051/0004-6361/201015880}, 529, 1

\bibitem[\protect\citeauthoryear{Coughlin \& Nixon}{Coughlin \&
  Nixon}{2015}]{coughlin2015-variability}
Coughlin E.~R.,  Nixon C.,  2015, \mn@doi [\apjl]
  {10.1088/2041-8205/808/1/L11}, 808, L11

\bibitem[\protect\citeauthoryear{Coughlin, Nixon, Begelman  \&
  Armitage}{Coughlin et~al.}{2016}]{coughlin2016-structure}
Coughlin E.~R.,  Nixon C.,  Begelman M.~C.,   Armitage P.~J.,  2016, \mn@doi
  [\mnras] {10.1093/mnras/stw770}, 17, 1

\bibitem[\protect\citeauthoryear{Curd \& Narayan}{Curd \&
  Narayan}{2018}]{curd2018}
Curd B.,  Narayan R.,  2018, \mn@doi [\mnras] {10.1093/mnras/sty3134}, 592, 565

\bibitem[\protect\citeauthoryear{Dai, McKinney  \& Miller}{Dai
  et~al.}{2015}]{dai2015}
Dai L.,  McKinney J.~C.,   Miller M.~C.,  2015, \mn@doi [\apj]
  {10.1088/2041-8205/812/2/L39}, 812, L39

\bibitem[\protect\citeauthoryear{Dai, McKinney, Roth, Ramirez-Ruiz  \&
  Miller}{Dai et~al.}{2018}]{dai2018}
Dai L.,  McKinney J.~C.,  Roth N.,  Ramirez-Ruiz E.,   Miller M.~C.,  2018,
  \mn@doi [\apj] {10.3847/2041-8213/aab429}, 859, L20

\bibitem[\protect\citeauthoryear{Draine}{Draine}{2011}]{draine2011}
Draine B.~T.,  2011, {Physics of the Interstellar and Intergalactic Medium}.
Princeton University Press

\bibitem[\protect\citeauthoryear{Evans \& Kochanek}{Evans \&
  Kochanek}{1989}]{evans1989}
Evans C.~R.,  Kochanek C.~S.,  1989, \mn@doi [\apj] {10.1086/185567}, 346, L13

\bibitem[\protect\citeauthoryear{Frank, King  \& Raine}{Frank
  et~al.}{2002}]{frank2002}
Frank J.,  King A.,   Raine D.,  2002, {Accretion Power in Astrophysics: Third
  Edition}.
Cambridge University Press, Cambridge

\bibitem[\protect\citeauthoryear{Gezari et~al.,}{Gezari
  et~al.}{2012}]{gezari2012}
Gezari S.,  et~al., 2012, \mn@doi [Nature] {10.1038/nature10990}, 485, 217

\bibitem[\protect\citeauthoryear{Guillochon \& McCourt}{Guillochon \&
  McCourt}{2017}]{guillochon2017-magnetic}
Guillochon J.,  McCourt M.,  2017, \mn@doi [\apj]
  {10.3847/2041-8213/834/2/L19}, 834, L19

\bibitem[\protect\citeauthoryear{Guillochon \& Ramirez-Ruiz}{Guillochon \&
  Ramirez-Ruiz}{2013}]{guillochon2013}
Guillochon J.,  Ramirez-Ruiz E.,  2013, \mn@doi [\apj]
  {10.1088/0004-637X/767/1/25}, 767, 25

\bibitem[\protect\citeauthoryear{Guillochon, Manukian  \&
  Ramirez-Ruiz}{Guillochon et~al.}{2014}]{guillochon2014-10jh}
Guillochon J.,  Manukian H.,   Ramirez-Ruiz E.,  2014, \mn@doi [\apj]
  {10.1088/0004-637X/783/1/23}, 783, 23

\bibitem[\protect\citeauthoryear{Hayasaki, Stone  \& Loeb}{Hayasaki
  et~al.}{2013}]{hayasaki2013}
Hayasaki K.,  Stone N.,   Loeb A.,  2013, \mn@doi [\mnras]
  {10.1093/mnras/stt871}, 434, 909

\bibitem[\protect\citeauthoryear{Hayasaki, Stone  \& Loeb}{Hayasaki
  et~al.}{2016}]{hayasaki2016-spin}
Hayasaki K.,  Stone N.,   Loeb A.,  2016, \mn@doi [\mnras]
  {10.1093/mnras/stw1387}, 461, 3760

\bibitem[\protect\citeauthoryear{Holoien et~al.,}{Holoien
  et~al.}{2019a}]{holoien2019}
Holoien T. W.~S.,  et~al., 2019a, preprint (\mn@eprint {arXiv} {1904.09293})

\bibitem[\protect\citeauthoryear{Holoien et~al.,}{Holoien
  et~al.}{2019b}]{holoien2019-18kh}
Holoien T. W.-S.,  et~al., 2019b, \mn@doi [\apj] {10.3847/1538-4357/ab2ae1},
  880, 120

\bibitem[\protect\citeauthoryear{Holoien et~al.,}{Holoien
  et~al.}{2020}]{holoien2020}
Holoien T. W.-S.,  et~al., 2020, \mn@doi [\apj] {10.3847/1538-4357/ab9f3d},
  898, 161

\bibitem[\protect\citeauthoryear{Hopkins}{Hopkins}{2015}]{hopkins2015}
Hopkins P.~F.,  2015, \mn@doi [\mnras] {10.1093/mnras/stv195}, 450, 53

\bibitem[\protect\citeauthoryear{Hung et~al.,}{Hung et~al.}{2020}]{hung2020}
Hung T.,  et~al., 2020, preprint (\mn@eprint {arXiv} {2003.09427})

\bibitem[\protect\citeauthoryear{Jiang, Guillochon  \& Loeb}{Jiang
  et~al.}{2016}]{jiang2016}
Jiang Y.-F.,  Guillochon J.,   Loeb A.,  2016, \mn@doi [\apj]
  {10.3847/0004-637X/830/2/125}, 830, 125

\bibitem[\protect\citeauthoryear{Kasen \& Ramirez-Ruiz}{Kasen \&
  Ramirez-Ruiz}{2010}]{kasen2010}
Kasen D.,  Ramirez-Ruiz E.,  2010, \mn@doi [\apj]
  {10.1088/0004-637X/714/1/155}, 714, 155

\bibitem[\protect\citeauthoryear{Kathirgamaraju, Duran  \&
  Giannios}{Kathirgamaraju et~al.}{2017}]{kathirgamaraju2017}
Kathirgamaraju A.,  Duran R.~B.,   Giannios D.,  2017, preprint (\mn@eprint
  {arXiv} {1701.07826})

\bibitem[\protect\citeauthoryear{Khabibullin, Sazonov  \& Sunyaev}{Khabibullin
  et~al.}{2013}]{khabibullin2014}
Khabibullin I.,  Sazonov S.,   Sunyaev R.,  2013, \mn@doi [\mnras]
  {10.1093/mnras/stt1889}, 437, 327

\bibitem[\protect\citeauthoryear{Kim, Park  \& Lee}{Kim et~al.}{1999}]{kim1999}
Kim S.~S.,  Park M.,   Lee H.~M.,  1999, \mn@doi [\apj] {10.1086/307394}, 519,
  647

\bibitem[\protect\citeauthoryear{Kochanek}{Kochanek}{1994}]{kochanek1994}
Kochanek C.~S.,  1994, \mn@doi [\apj] {10.1086/173745}, 422, 508

\bibitem[\protect\citeauthoryear{Leloudas et~al.,}{Leloudas
  et~al.}{2019}]{leloudas2019}
Leloudas G.,  et~al., 2019, \mn@doi [\apj] {10.3847/1538-4357/ab5792}, 887, 218

\bibitem[\protect\citeauthoryear{Levermore \& Pomraning}{Levermore \&
  Pomraning}{1981}]{levermore1981}
Levermore C.~D.,  Pomraning G.~C.,  1981, \mn@doi [\apj] {10.1086/159157}, 248,
  321

\bibitem[\protect\citeauthoryear{Liptai, Price, Mandel  \& Lodato}{Liptai
  et~al.}{2019}]{liptai2019-spin}
Liptai D.,  Price D.~J.,  Mandel I.,   Lodato G.,  2019, preprint, 12, 1
  (\mn@eprint {arXiv} {1910.10154})

\bibitem[\protect\citeauthoryear{Lodato, King  \& Pringle}{Lodato
  et~al.}{2009}]{lodato2009}
Lodato G.,  King a.~R.,   Pringle J.~E.,  2009, \mn@doi [\mnras]
  {10.1111/j.1365-2966.2008.14049.x}, 392, 332

\bibitem[\protect\citeauthoryear{Lu \& Bonnerot}{Lu \& Bonnerot}{2020}]{lu2020}
Lu W.,  Bonnerot C.,  2020, \mn@doi [\mnras] {10.1093/mnras/stz3405}, 492, 686

\bibitem[\protect\citeauthoryear{Matsumoto \& Piran}{Matsumoto \&
  Piran}{2020}]{matsumoto2020}
Matsumoto T.,  Piran T.,  2020, preprint, 9, 1 (\mn@eprint {arXiv}
  {2009.01240})

\bibitem[\protect\citeauthoryear{Metzger \& Stone}{Metzger \&
  Stone}{2016}]{metzger2016}
Metzger B.~D.,  Stone N.~C.,  2016, \mn@doi [\mnras] {10.1093/mnras/stw1394},
  461, 948

\bibitem[\protect\citeauthoryear{Mihalas \& Mihalas}{Mihalas \&
  Mihalas}{1984}]{mihalas1984}
Mihalas D.,  Mihalas B.,  1984, {Foundations of radiation hydrodynamics}.
Oxford Univ. Press, Oxford

\bibitem[\protect\citeauthoryear{Nicholl et~al.,}{Nicholl
  et~al.}{2019}]{nicholl2019}
Nicholl M.,  et~al., 2019, \mn@doi [\mnras] {10.1093/mnras/stz1837}, 488, 1878

\bibitem[\protect\citeauthoryear{Nicholl et~al.,}{Nicholl
  et~al.}{2020}]{nicholl2020}
Nicholl M.,  et~al., 2020, preprint, 20, 1 (\mn@eprint {arXiv} {2006.02454})

\bibitem[\protect\citeauthoryear{Phinney}{Phinney}{1989}]{phinney1989}
Phinney E.~S.,  1989, in Morris M.,  ed.,  Proc. IAU Symposium Vol. 136, The
  Center of the Galaxy. Kluwer, Dordrecht, p.~543

\bibitem[\protect\citeauthoryear{Piran, Svirski, Krolik, Cheng  \&
  Shiokawa}{Piran et~al.}{2015}]{piran2015-circ}
Piran T.,  Svirski G.,  Krolik J.,  Cheng R.~M.,   Shiokawa H.,  2015, \mn@doi
  [\apj] {10.1088/0004-637X/806/2/164}, 806, 164

\bibitem[\protect\citeauthoryear{Press, Teukolsky, Vetterling  \&
  Flannery}{Press et~al.}{1992}]{press1992}
Press W.,  Teukolsky S.,  Vetterling W.,   Flannery B.,  1992, {Numerical
  recipes in FORTRAN. The art of scientific computing}.
Cambridge University Press

\bibitem[\protect\citeauthoryear{Price}{Price}{2007}]{price2007}
Price D.~J.,  2007, \mn@doi [PASA] {10.1071/AS07022}, 24, 159

\bibitem[\protect\citeauthoryear{Ramirez-Ruiz \& Rosswog}{Ramirez-Ruiz \&
  Rosswog}{2009}]{ramirez-ruiz2009}
Ramirez-Ruiz E.,  Rosswog S.,  2009, \mn@doi [\apj]
  {10.1088/0004-637X/697/2/L77}, 697, L77

\bibitem[\protect\citeauthoryear{Rees}{Rees}{1988}]{rees1988}
Rees M.~J.,  1988, \mn@doi [Nature] {10.1038/333523a0}, 333, 523

\bibitem[\protect\citeauthoryear{Roth, Kasen, Guillochon  \& Ramirez-ruiz}{Roth
  et~al.}{2016}]{roth2016}
Roth N.,  Kasen D.,  Guillochon J.,   Ramirez-ruiz E.,  2016, \mn@doi [\apj]
  {10.3847/0004-637X/827/1/3}, 827, 1

\bibitem[\protect\citeauthoryear{Rybicki \& Lightman}{Rybicki \&
  Lightman}{1979}]{rybicki1979}
Rybicki G.,  Lightman A.,  1979, {Radiative processes in astrophysics}.
New York, Wiley-Interscience

\bibitem[\protect\citeauthoryear{Sadowski, Tejeda, Gafton, Rosswog  \&
  Abarca}{Sadowski et~al.}{2016}]{sadowski2016}
Sadowski A.,  Tejeda E.,  Gafton E.,  Rosswog S.,   Abarca D.,  2016, \mn@doi
  [\mnras] {10.1093/mnras/stw589}, 458, 4250

\bibitem[\protect\citeauthoryear{Shiokawa, Krolik, Cheng, Piran  \&
  Noble}{Shiokawa et~al.}{2015}]{shiokawa2015}
Shiokawa H.,  Krolik J.~H.,  Cheng R.~M.,  Piran T.,   Noble S.~C.,  2015,
  \mn@doi [\apj] {10.1088/0004-637X/804/2/85}, 804, 85

\bibitem[\protect\citeauthoryear{Short et~al.,}{Short et~al.}{2020}]{short2020}
Short P.,  et~al., 2020, preprint, 19, 1 (\mn@eprint {arXiv} {2003.05470})

\bibitem[\protect\citeauthoryear{Steinberg, Coughlin, Stone  \&
  Metzger}{Steinberg et~al.}{2019}]{steinberg2019}
Steinberg E.,  Coughlin E.~R.,  Stone N.~C.,   Metzger B.~D.,  2019, \mn@doi
  [Mon. Not. R. Astron. Soc. Lett.] {10.1093/mnrasl/slz048}, 485, L146

\bibitem[\protect\citeauthoryear{Stone \& Loeb}{Stone \&
  Loeb}{2012}]{stone2012}
Stone N.,  Loeb A.,  2012, \mn@doi [Phys. Rev. Lett.]
  {10.1103/PhysRevLett.108.061302}, 108, 061302

\bibitem[\protect\citeauthoryear{Stone, Sari  \& Loeb}{Stone
  et~al.}{2013}]{stone2013}
Stone N.,  Sari R.,   Loeb A.,  2013, \mn@doi [\mnras] {10.1093/mnras/stt1270},
  435, 1809

\bibitem[\protect\citeauthoryear{Turner \& Stone}{Turner \&
  Stone}{2001}]{turner2001}
Turner N.~J.,  Stone J.~M.,  2001, \mn@doi [\apjs] {10.1086/321779}, 135, 95

\bibitem[\protect\citeauthoryear{Whitehouse \& Bate}{Whitehouse \&
  Bate}{2004}]{whitehouse2004}
Whitehouse S.~C.,  Bate M.~R.,  2004, \mn@doi [\mnras]
  {10.1111/j.1365-2966.2004.08131.x}, 353, 1078

\bibitem[\protect\citeauthoryear{van Velzen et~al.,}{van Velzen
  et~al.}{2019}]{van_velzen2019}
van Velzen S.,  et~al., 2019, \mn@doi [\apj] {10.3847/1538-4357/aafe0c}, 872,
  198

\makeatother
\end{thebibliography}




\bsp	
\label{lastpage}


\end{document}